\documentclass[12pt]{article}
\usepackage{amssymb}
\usepackage{amsmath}
\usepackage{amstext}
\usepackage{graphicx,epsfig}
\usepackage{epsfig}
\usepackage{verbatim} 
\usepackage{fancybox}
\usepackage{color}
\usepackage{ulem}
\usepackage{enumitem}
\usepackage{subfigure}
\usepackage{bbm}
\usepackage{parskip}
\usepackage{dsfont}
\usepackage[numbers,sort&compress]{natbib}
\usepackage[dvipsnames]{xcolor}
\usepackage{cancel}
\usepackage[numbers,sort&compress]{natbib}

\newcommand{\Comment}[1]{{}}
\definecolor{darkblue}{rgb}{0.95,0.05,0.05}
\definecolor{reddish}{rgb}{0.95, 0.05, 0.05}
\usepackage[linktocpage=true]{hyperref}
\hypersetup{
colorlinks=true,
citecolor=darkblue,
linkcolor=reddish,
urlcolor=darkblue,
pdfauthor={},
pdftitle={},
pdfsubject={}
}

\def\({\left(}
\def\){\right)}

\newcommand{\beq}{\begin{equation}}
\newcommand{\eeq}{\end{equation}}
\newcommand{\be}{\begin{equation}}
\newcommand{\ee}{\end{equation}}
\newcommand{\bea}{\begin{align}}
\newcommand{\eea}{\end{align}}

\def\gsim{ \lower .75ex \hbox{$\sim$} \llap{\raise .27ex \hbox{$>$}} }
\def\lsim{ \lower .75ex \hbox{$\sim$} \llap{\raise .27ex \hbox{$<$}} }

\def\beq{\begin{eqnarray}}
\def\eeq{\end{eqnarray}}

\def\mpl{M_{\rm Pl}}

\def\p{{\cal P}}

\def\L*{{\cal L}_*}
\def\L{\mathcal{L}}
\def\({\left(}
\def\){\right)}

\def\p{\partial}

\def\p{\partial}

\def\<{\langle}
\def\>{\rangle}

\def\xyma{\xymatrix@M.7em}
\def\xymas{\xymatrix@M.1em}
\newcommand{\ba}{\begin{eqnarray}}
\newcommand{\ea}{\end{eqnarray}}

\usepackage{tikz}
\usetikzlibrary{decorations}
\pgfdeclaredecoration{complete sines}{initial}
{
    \state{initial}[
        width=+0pt,
        next state=upsine,
        persistent precomputation={\pgfmathsetmacro\matchinglength{
            \pgfdecoratedinputsegmentlength / int(\pgfdecoratedinputsegmentlength/\pgfdecorationsegmentlength)}
            \setlength{\pgfdecorationsegmentlength}{\matchinglength pt}
        }] {}
    \state{upsine}[width=\pgfdecorationsegmentlength,next state=downsine]{
        \pgfpathsine{\pgfpoint{0.25\pgfdecorationsegmentlength}{0.5\pgfdecorationsegmentamplitude}}
        \pgfpathcosine{\pgfpoint{0.25\pgfdecorationsegmentlength}{-0.5\pgfdecorationsegmentamplitude}}
    }
    \state{downsine}[width=\pgfdecorationsegmentlength,next state=upsine]{
        \pgfpathsine{\pgfpoint{0.25\pgfdecorationsegmentlength}{-0.5\pgfdecorationsegmentamplitude}}
        \pgfpathcosine{\pgfpoint{0.25\pgfdecorationsegmentlength}{0.5\pgfdecorationsegmentamplitude}}
}
    \state{final}{}
}

\title{}
\author{}

\numberwithin{equation}{section}

\begin{document}
%
\renewcommand{\thefootnote}{\fnsymbol{footnote}}
~
%
%
\begin{center}
{\LARGE \bf{Relaxing the Cosmological Constant:}} 
\vskip 0.2cm
{\LARGE \bf{a Proof of Concept}}
\end{center} 
 \vspace{1truecm}
\thispagestyle{empty} \centerline{
{\Large  {Lasma Alberte,${}^{\rm a}$}}
{\Large  {Paolo Creminelli,${}^{\rm b}$}}
{\Large {Andrei Khmelnitsky,${}^{\rm b}$}}
}
\vskip 0.2cm
\centerline{
{\Large {David Pirtskhalava,${}^{\rm c}$}}
{\Large {Enrico Trincherini${~}^{\rm d}$}}
}

\vspace{.5cm}

\centerline{{${}^{\rm a}$\it SISSA, Via Bonomea 265, 34136 Trieste, Italy}}
\centerline{{\it INFN -- Sezione di Trieste, Via Valerio 2, 34127 Trieste, Italy}}
\vskip 0.1cm
\centerline{{${}^{\rm b}$\it Abdus Salam International Centre for Theoretical Physics (ICTP)}}
\centerline{{\it Strada Costiera 11, 34151, Trieste, Italy}}
\vskip 0.1cm
\centerline{{${}^{\rm c}$\it Institute of Physics, \'Ecole Polytechnique F\'ed\'erale de Lausanne}}
\centerline{{\it CH-1015, Lausanne, Switzerland}}
\vskip 0.1cm
\centerline{{${}^{\rm d}$\it Scuola Normale Superiore, Piazza dei Cavalieri 7, 56126, Pisa, Italy}}
\centerline{{\it INFN -- Sezione di Pisa, 56200, Pisa, Italy}}

 \vspace{1cm}
\begin{abstract}
\noindent
We propose a technically natural scenario whereby an initially large cosmological constant (c.c.)~is relaxed down to the observed value due to the dynamics of a scalar evolving on a very shallow potential. The model crucially relies on a sector that violates the null energy condition (NEC) and gets activated only when the Hubble rate becomes sufficiently small --- of the order of the present one. As a result of NEC violation, this low-energy universe evolves into inflation, followed by reheating and the standard Big Bang cosmology. The symmetries of the theory force the c.c.~to be the same before and after the NEC-violating phase, so that a late-time observer sees an effective c.c.~of the correct magnitude. Importantly, our model allows neither for eternal inflation nor for a set of possible values of dark energy, the latter fixed by the parameters of the theory. 
\end{abstract}

\newpage

\setcounter{tocdepth}{2}
\tableofcontents
\newpage
\renewcommand*{\thefootnote}{\arabic{footnote}}
\setcounter{footnote}{0}

\section{Introduction}
The cosmological constant problem is arguably the biggest conundrum in physics today. It is fair to say that the ``anthropic'' explanation  \cite{Weinberg:1988cp} is at the moment the most compelling, even if the lack of a concrete way of testing it leaves a lot of room for personal taste and heated discussions. Anthropic explanations have the danger of ``premature application'', quoting S.~Dimopoulos, and the purpose of this paper is to try to find an alternative explanation for the smallness of the c.c. We are going to discuss a ``historical'' explanation in which the universe dynamically evolves towards a state with a small vacuum energy and we are heavily inspired by Abbott's original relaxation mechanism \cite{Abbott:1984qf} and more recent ideas about the relaxation of the electroweak hierarchy \cite{Graham:2015cka} (for earlier studies, see \cite{Dvali:2003br,Dvali:2004tma}).~The idea of the model is simple (while its actual implementation is not). A scalar, $\phi_1$, moving on a potential with a slight negative tilt slowly scans the value of the vacuum energy, starting from large and positive values.\footnote{Since we will be considering time-dependent solutions, the relaxation evades Weinberg's theorem \cite{Weinberg:1988cp}.} To avoid large quantum fluctuations and thus eternal inflation with all its disturbing consequences, we will need this field to be a ghost condensate  \cite{ArkaniHamed:2003uy} (Section \ref{sec:relaxing}). As the field progresses, the Hubble friction becomes smaller and smaller and changes the dynamics of relaxation; this triggers a phase transition in a second sector of the theory, described by a scalar $\phi_2$ (Section~\ref{sec:bound}). Crucially, this second sector needs to violate the Null Energy Condition: this is mandatory since any relaxation mechanism has to probe the small energies associated with the observed c.c.~before the standard cosmology takes place  \cite{Steinhardt:2006bf}. We are going to present two models which violate the NEC and give rise to the observable universe: one based on another ghost condensate (Section \ref{sec:slownec}) --- as promised the implementation of the idea is not simple --- and the second on a Galileon theory (Section \ref{sec:fastnec}). In both cases a symmetry forces the c.c~to be the same before and after the NEC violating phase, so that the observed value of the c.c.~is the one relaxed early on. The transfer of energy from the NEC violating sector to standard matter is realized with the aid of a waterfall field $\chi$ (Section \ref{sec:reheating}), similarly to what happens in hybrid inflation \cite{Linde:1993cn}. 

\begin{figure}[h!]
\center
\includegraphics[width=16cm]{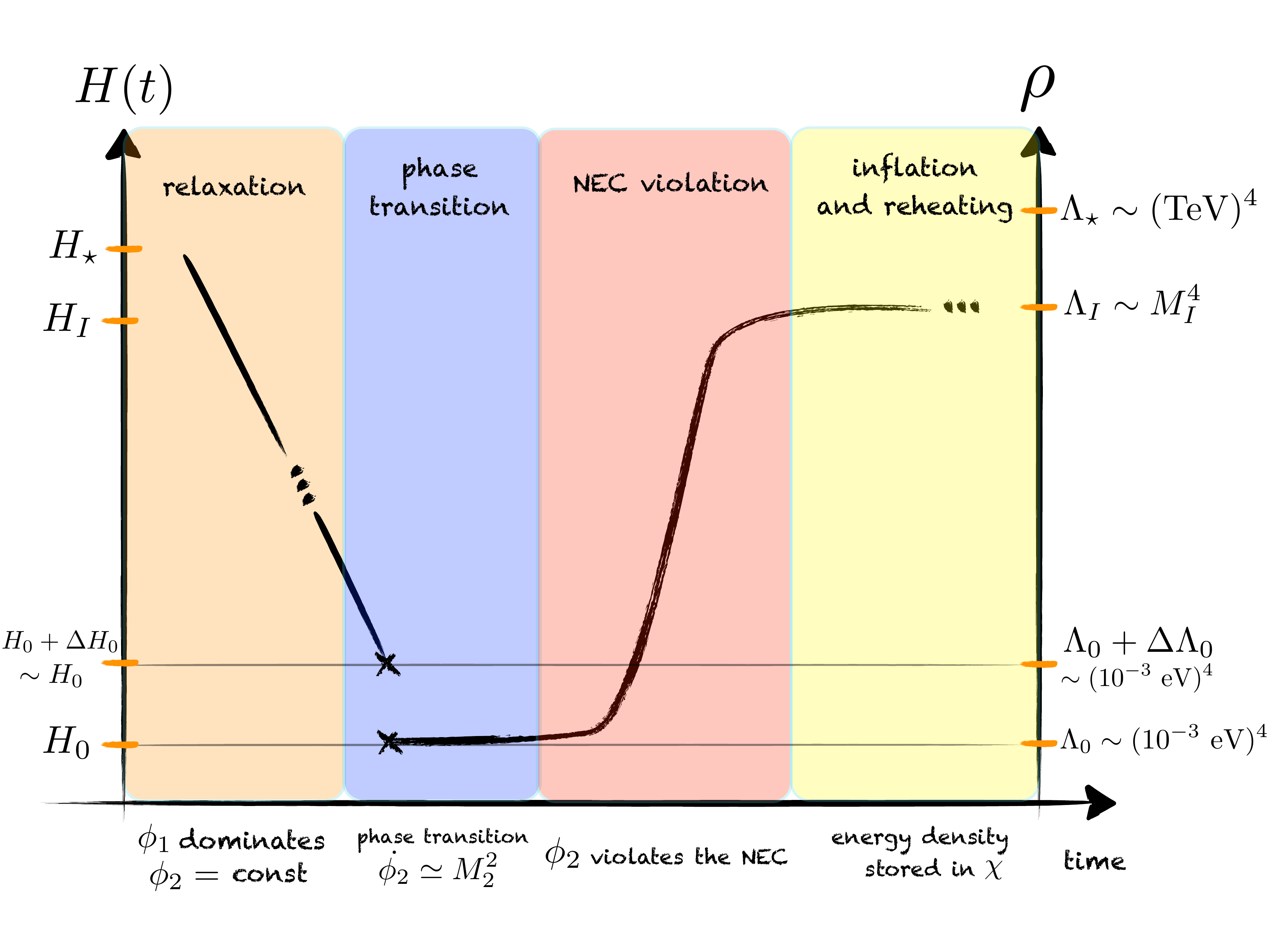}%
\caption{A sketch of the universe's expansion rate as a function of time in our model. More details on each of the four stages (relaxation, phase transition in the NEC-violating sector, violation of the NEC, inflation/reheating) as well as on the notation used are provided in the following sections. }
\label{fig:intro}
\end{figure}

It is important to stress that our model does not allow for a set of possible values of the observed vacuum energy: the universe violates the NEC and reheats when a particular value of the c.c.~is reached. In this aspect the model is at variance with \cite{Abbott:1984qf,Steinhardt:2006bf,Graham:2015cka}. Once different values for the c.c.~or the electroweak scale are possible the Pandora's box is open: one is forced to address the measure problem in the landscape and to ponder on the anthropic weight, completely defeating the purpose. 

Notice that a dynamical solution of the c.c.~problem requires a full description of the whole cosmological evolution, since the relaxation happens {\it before} the standard cosmology. In particular, the NEC violating phase must be followed by a period of inflation (or some alternative thereof) to give rise to the scalar perturbations we observe. The resulting model is clearly complicated (see Fig.~\ref{fig:intro} for a schematic sketch) and involves very different energy scales and vacuum expectation values which, although technically natural, make the picture aesthetically unpleasant. It is difficult to believe the universe really works in this way. This may be due to the authors' lack of imagination and more compelling models may be found. 

We see at least two strong motivations to push further the idea of relaxing the c.c. First, there is a huge experimental activity to test dark energy and its phenomenological differences from a pure vacuum energy. The only {\it raison d'\^etre} of dark energy is the cosmological constant problem. However, ironically, almost none of the discussed models addresses this problem in any way (and some introduce additional fine tunings on top of the c.c.~one). Notable exceptions are, for example, ``global'' modifications of gravity \cite{Gabadadze:2014rwa,Gabadadze:2015goa,Kaloper:2015jra} and ``degravitation'' models \cite{ArkaniHamed:2002fu,Dvali:2007kt,Derham:2010tw}. While there is a strong activity in constraining generic models of dark energy, there is no reason to expect their phenomenology has anything to do with the physics which solves the c.c.~problem (if any). In our scenario, the two sectors (the one relaxing the vacuum energy and the one violating the NEC) are also dark energy components nowadays, but now indeed related to the c.c.~problem! The present dark energy is related, albeit in a model-dependent way, to the violation of the NEC in the past and does not reduce simply to a small vacuum energy. The second motivation is fully theoretical. A dynamical relaxation of the c.c.~needs a subsequent violation of the NEC. It is not yet clear whether some general UV obstructions to building NEC violating theories exist. Stopping the exploration in this direction would be a clear premature application of the anthropic principle.


\section{\label{sec:relaxing}Relaxing the c.c.~without eternal inflation}

The mechanism responsible for relaxing the cosmological constant ought to satisfy two basic requirements. First, if that mechanism is to be free from fine tuning, it must be stable under order-unity variations of the initial vacuum energy. And second, it has to be dominated by classical dynamics in order to bypass the standard problems with defining the cosmological probability measures.  We will see that these two requirements alone significantly constrain the ways in which the relaxation of the cosmological constant can be realized. 

Perhaps the simplest realization that complies with the above conditions is provided by an approximately shift-symmetric scalar $\phi_1$, governed by the following low-energy effective action
\be
\label{ghost}
S = \int d^4 x \sqrt{-g} \bigg[M_1^4 P_1\(X_1\)+\lambda_1^3 \phi_1 -\Lambda_\star  + \dots\bigg ].
\ee
Here, $\Lambda_\star=3\mpl^2H_\star^2$ is the cosmological constant we wish to relax (which we assume to be positive), $\lambda_1$ is set by some small scale (with `small' quantified shortly), and $P_1$ is a generic function of 
\be
X_1 \equiv -\frac{g^{\mu\nu}\p_\mu\phi_1 \p_\nu\phi_1}{M_1^4}~.
\ee
Furthermore, by the ellipses we denote all other operators in the effective theory, suppressed by inverse powers of the cutoff $M_1$ and/or the shift symmetry-breaking spurion\footnote{In the absence of gravity, $\lambda_1$ does \textit{not} break the scalar's shift symmetry, meaning that the symmetry-breaking terms are naturally suppressed both by $\lambda_1$ and by further powers of the Planck mass.} $\lambda_1$. 

Suppose now that the scalar is canonical, $P_1 = X_1/2$, and that it starts slowly rolling down the linear potential from a generic point where the effective dark energy density (that also includes the contribution from the scalar's potential energy) is of the order of $\Lambda_\star$. In the course of the evolution, the cosmological constant adiabatically decreases. (Since the universe is inflating, all other sources of energy will be quickly diluted away.) Slow roll ($|\dot H|\ll H^2$) requires a sufficiently flat potential
\be
\label{c1}
\lambda_1^3 \ll \mpl H^2  \qquad (\text{slow roll})~,
\ee
while the condition that the dynamics be classical reads
\beq
\label{c2}
\frac{\dot{\phi_1}}{H}\gg  H \quad\Rightarrow\quad \lambda_1^3 \gg H^3  \qquad (\text{classical evolution})~.
\eeq
Requiring \eqref{c1} and \eqref{c2} to be marginally satisfied respectively for the present value of the expansion rate (when $H\simeq H_0\sim 10^{-33}~ \rm{eV}$) and at the initial stages of relaxation (when $H\simeq H_\star$) yields an upper bound $H_\star \ll (\mpl H_0^2)^{1/3}$. This corresponds to the following maximal value of the relaxed energy density 
\beq
\label{lambdamax}
\Lambda_{\rm max} \sim \mpl^{8/3} H_0^{4/3} \sim (10~ \text{MeV})^4~.
\eeq  
Thus, a slowly rolling canonical scalar can not relax a cosmological constant larger than $\sim (10~ \text{MeV})^4$ if it is to conform to purely classical dynamics all along. 
\begin{figure}[t!]
	\center
	\includegraphics[width=12cm]{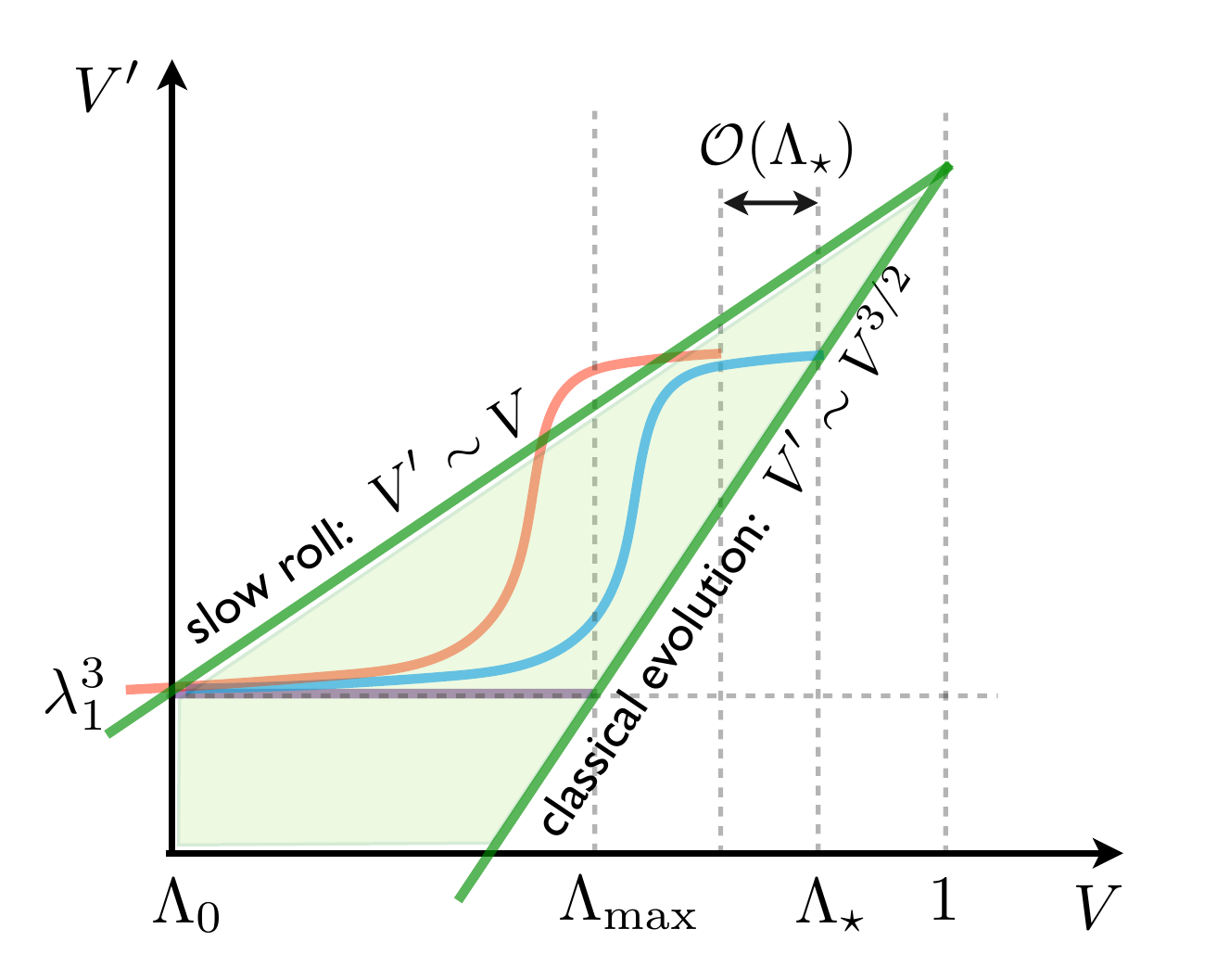}%
	\caption{The slope of the potential as a function of the effective dark energy $V=-\lambda_1^3\phi_1+\Lambda_\star$. The green region corresponds to the allowed window, compatible with both the slow-roll dynamics and classical evolution. Adjusting the slope so as to make it optimally compatible with both requirements (blue curve) inevitably entails fine tuning. A change in the initial c.c.~of the order of $\Lambda_\star$ leads to a breakdown of the slow-roll regime before the cosmological constant relaxes to sufficiently low values (red curve). The case corresponding to a constant slope is depicted by the horizontal purple line. The figure is not to scale and we have set $\mpl$ to one.}
	\label{fig:sr}
\end{figure}

The last conclusion draws heavily upon assuming a constant slope of the scalar potential. One can in principle give up this requirement, allowing for a slope that changes adiabatically in the course of the evolution, in a way that is optimally compatible with both conditions  \eqref{c1} and \eqref{c2}. One could imagine a potential which is steep at the early stages of relaxation (when $H\sim H_\star$), while becoming flat at  times when the Hubble rate drops down to $\sim H_0$.   Such a potential can be made compatible with both slow roll and classical evolution all along the relaxation trajectory by tuning the tilt $V'$ to lie in between $V/\mpl$ and $(V^{1/2}/\mpl)^3$ at any particular moment of time. The corresponding situation is depicted in Fig.~\ref{fig:sr}. However, it necessarily entails fine-tuning: an order-unity variation of the initial cosmological constant (without changing the potential and the initial conditions) would  result in a breakdown of one of the above two conditions way before the effective cosmological constant drops down to the desired value. In other words, the potential has to `know' when $H$ becomes small, and this brings back the usual fine tuning of the c.c. 

The constraint \eqref{c2} that arises from requiring classical evolution of $\phi_1$ can be made milder\footnote{For example, this happens in {\it k-inflation} models \cite{ArmendarizPicon:1999rj}.} and even removed altogether if the background dynamics is in a different, \textit{ghost-condensate} regime \cite{ArkaniHamed:2003uy}.\footnote{We refer to this regime as `ghost condensate', even though $\phi_1$ does not necessarily have to describe a ghost on its Poincar\'e-invariant vacuum ($\phi_1=\rm{const}$), which may or may not be connected to $\dot\phi_1\neq 0$ vacua within the same low-energy EFT (of course, the Poincar\'e-invariant vacuum only exists in the limit $\lambda_1\to 0$). } This is described by a particular attractor solution with a constant velocity that corresponds to a minimum of the function $P_1$:
\beq
\label{x1}
X_1 = \frac{\dot{\phi_1^2}}{M_1^4} = 1~.
\eeq  
For a non-vanishing tilt of the potential, this solution is slightly perturbed by a homogeneous mode $\pi_1 \equiv \phi_1 - M_1^2 t$~, whose velocity is driven to the following terminal value by the expansion of the universe\footnote{We assume the canonical normalization for $\pi_1$, which corresponds to $P''_1(1)=1/4$ \cite{ArkaniHamed:2003uy,ArkaniHamed:2003uz}.}
\be
\label{terminalvelocity}
\dot\pi_1 \simeq \frac{\lambda_1^3}{3H}~.
\ee
Imposing that $\pi$ be a small perturbation then yields
$\lambda_1^3 \ll 3 H M_1^2$.

For our purposes, the virtue of the ghost condensate is that the field always evolves classically as long as the cutoff of the theory is well above the Hubble scale (that is, as far as the low-energy EFT is valid). In particular, the relative quantum versus classical variation of $\phi_1$ over an e-fold reads \cite{ArkaniHamed:2003uz} 
\begin{eqnarray}
\label{spectrum}
\frac{\(\delta \phi_1\)_{\rm quant}}{\(\delta \phi_1\)_{\rm class}}\sim \(\frac{H}{M_1} \)^{5/4}\,.
\end{eqnarray}
The peculiar expression  for the spectrum of the scalar's quantum  fluctuations arises from the fact that their gradient energy comes from higher derivative operators in the effective theory, while the quadratic in momentum contribution vanishes at the leading order in $\epsilon\equiv -\dot H/H^2$. This results in a dispersion relation of the form $\omega^2 \sim k^4/M_1^2$ around Hubble frequencies, leading to \eqref{spectrum}. Most importantly for our purposes, the amplitude of quantum fluctuations is independent of the tilt of the potential and remains small even in the limit $\lambda_1\to 0$. (Although we stick here to the ghost-condensate for concreteness, one can consider other models that keep a classical motion in the $\lambda_1 \to 0 $ limit, for example based on Galilean symmetry \cite{Kobayashi:2010cm}.)

The very same higher-derivative operators that determine the spectrum of short-wavelength perturbations of $\phi_1$ also induce a Jeans-like instability for long-wavelength modes, once the effects of mixing with gravity are taken into account \cite{ArkaniHamed:2003uy}. Requiring the characteristic time scale of this instability to be longer than the current Hubble time strongly constrains the cutoff of the theory \cite{ArkaniHamed:2003uy}:
\be
\label{jeans}
M_1^3 < \mpl^2 H_0\sim (10~ \text{MeV})^3~.
\ee
Using this constraint and imposing that the relaxation proceeds within the regime of validity of the low-energy effective field theory (\textit{i.e.}~$M_1>H_\star$) yields the following upper bound on the magnitude of the relaxed cosmological constant: 
$
\Lambda_\star ~\lsim~ \mpl^{10/3} H_0^{2/3}\sim \(10^5 ~\text{TeV}\)^4.$
We will see in what follows that this estimate is too optimistic: the structure of the model imposes $M_1~\lesssim ~10^{-3}~\rm{eV}$, which is well compatible with the bound \eqref{jeans} imposed by stability. The validity of the low energy EFT thus requires $\Lambda_\star~\lsim~ (1~\rm{TeV})^4~$.

The ghost condensate entails no constraint on the tilt of the potential from the requirement of classical evolution. However, the upper bound from imposing a quasi-stationary relaxation, $\epsilon \ll 1$, is still there, and it reads: 
\beq
\label{gcslowroll}
\epsilon_0\equiv \(-\frac{\dot H}{H^2}\)_{H=H_0} \simeq \frac{\lambda_1^3 M_1^2}{6\mpl^2 H_0^3} \ll 1~.
\eeq	
This constraint can be always satisfied by taking $\lambda_1$ small enough. Moreover, using \eqref{gcslowroll}, the correction to the $\phi_1$ velocity \eqref{terminalvelocity} can be expressed through $\epsilon$ as 
\be
\label{pi0}
\(\frac{\dot\pi_1}{M_1^2}\)_{H=H_0} \simeq 2\frac{\mpl^2 H_0^2}{M_1^4}\epsilon_0~. 
\ee

Below, we will consider a scenario where the quasi-stationary evolution of the background breaks down (\textit{i.e.} $\epsilon$ becomes greater than one) not much later than when the Hubble rate drops to $H\sim H_0$. Eq. \eqref{pi0} then shows, that 
for $M_1^4\sim \mpl^2 H_0^2$ --- the value of the EFT cutoff we will be particularly interested in --- the background starts to deviate by order one from the ghost condensate solution \eqref{x1} around the same time. 

\section{\label{sec:bound}Phase transition to NEC violation}
\subsection{A bound on the trigger} 

After having relaxed the large cosmological constant $\Lambda_\star$, the universe finds itself in an empty state with tiny curvature. To turn this into a realistic scenario, one has to specify a NEC-violating mechanism responsible for creating energy density that corresponds to at least the lowest possible reheating temperature compatible with Big Bang nucleosynthesis, $T_{\text{reh}}~\gsim~ 5\text{ MeV}$ \cite{Kawasaki:2000en,Hannestad:2004px}. 
It is crucial that the dynamics that eventually leads to reheating get activated only when the relaxation has reached a certain stage: this will fix the observed value of the c.c. Notice that the relaxation mechanism guarantees the smallness of the vacuum energy in the low-temperature vacuum (there is no thermal plasma during the relaxation): after reheating the thermal history of the universe will include various phase transitions with corresponding jumps in the vacuum energy; however the c.c.~we observe today will be small provided the final vacuum is the same as the one we relaxed early on (or is related to it by a symmetry).

The violation of the NEC will be achieved by a field $\phi_2$ which, as we will see, has an action similar to \eqref{ghost} in most of the field space, with a cutoff $M_2$. We envision a sort of phase transition for $\phi_2$ triggered by the relaxation field $\phi_1$: at early times the field $\phi_2$ is at rest (or this degree of freedom may even not exist early on) and only later, once a certain value of the c.c.~is reached, a phase transition occurs to the NEC-violating regime.
Regardless of the details of the phase transition, it is clear that it will induce a change in the vacuum energy of the order of $\sim M_2^4$ and this imposes a severe constraint on the model. Since this jump occurs after the relaxation, it must be small compared to the observed c.c.~to avoid reintroducing fine tuning. Therefore, the NEC-violating sector has to be characterized by a low cutoff \footnote{To trust the low energy EFT of $\phi_2$, at no point during the relaxation process should the Hubble rate exceed this scale. This would impose an upper bound, $H_{\star}~\lsim~ M_2$, on the curvature of the universe, corresponding to a maximal value of the cosmological constant $\Lambda_{\star}\lesssim (1\text{ TeV})^4$.
This bound is not very robust, since there is nothing wrong if at the beginning of the relaxation one is sensitive to the UV completion of the $\phi_2$ sector. However a very similar bound will be derived below using constraints on the relaxation sector.}
\be\label{m2_max}
M_2~\lsim~ \Lambda_0^{1/4}\sim 10^{-3}\text{ eV}\;.
\ee


For the scenario at hand to work, it is obviously important that the effective c.c. has not been reduced significantly by the dynamics of $\phi_1$ taking place from the onset of the NEC-violating phase up to the present time. As we will see, this leads to significant bounds on the sector of our model responsible for violating the NEC.

\subsection{\label{sec:trigger}Dynamics of the transition and a bound on $\Lambda_\star$}

We have remarked above that the NEC-violating dynamics of $\phi_2$ should be triggered when the Hubble rate drops down to values of order $H_0$.
In principle, this can happen in a few different ways.
One could imagine that the effective action of $\phi_2$ is directly sensitive to the Hubble rate, for example as a consequence of integrating out a degree of freedom $\sigma$ with mass fixed by the curvature through a non-minimal coupling $R \sigma^2$.
Another possibility --- more along the lines of Refs. \cite{Abbott:1984qf,Graham:2015cka} --- to make the system sensitive to the varying Hubble rate, is to invoke strong dynamics. Unfortunately, neither of these possibilities work for our purposes. In the first case, the mass of $\sigma$ will receive radiative corrections at least of order $\Lambda_\star/\mpl^2 \sim H_\star^2$ so that, barring fine-tuning, the dependence on $H$ is irrelevant for $H \ll H_\star$. 
In the second case, one can invoke another sector that confines at energies around $H_0$ and affects the dynamics of $\phi_2$ in some way. However, the energy density that such a strongly-coupled sector can store is at most of order $H_0^4$ --- much less than the characteristic energy density of the NEC-violating sector of our model $M_2^4\sim \mpl^2 H_0^2$. This makes it practically impossible for the confining phase transition to influence the $\phi_2$ dynamics in any significant way.     

The other route --- the one we will stick to below --- to encode information about the background evolution into the EFT of $\phi_2$ is to couple it directly to the scanning scalar $\phi_1$ through some Lagrangian term
\beq
\label{mixing}
S\supset \int d^4x \sqrt{-g} M_2^4 P(X_1,X_2)~.
\eeq
The deviation $\pi_1$ of $\phi_1$ from the exact ghost condensate solution \eqref{x1} is negligible at early times while $X_2$ is assumed to vanish at that stage. However, $\pi_1$ grows with the relaxation of the cosmological constant as a result of the reduced Hubble friction: \textit{the system is thus naturally sensitive to the zero of the vacuum energy.} In particular, there is a significant change in the interactions described by \eqref{mixing} around the time when $\dot\pi_1/M_1^2$ becomes of order one.
We will assume, that this causes the dynamics of $\phi_1$ and $\phi_2$ to change qualitatively --- the latter scalar resetting after the phase transition onto its NEC-violating trajectory.  The observed value of the c.c.~is therefore fixed, using Eq.~\eqref{terminalvelocity}, in terms of the parameters of the model as
\be
\label{lambdavalue}
\Lambda_0 \sim \frac{\lambda_1^6 \mpl^2}{M_1^4} \;.
\ee
Depending on how the behaviour of the scanning field $\phi_1$ changes at the phase transition, one has very different constraints on the dynamics of the NEC-violating sector.

The first possibility is that $\phi_1$ gets stabilized in a trivial vacuum, $\phi_1 = \text{const}$, after the effective field theory for the scanning field breaks down at $\dot\pi_1/M_1^2 \sim 1$. The scanning of the cosmological constant therefore terminates at the value $\Lambda \sim \Lambda_0$ and there is essentially unlimited time for the NEC violation to proceed. We describe the corresponding scenario with slow NEC violation in Section~\ref{sec:slownec}. 
One expects the new vacuum of $\phi_1$ to have an energy density that differs from that in the rolling state by $\sim M_1^4$. In order for this change in the effective cosmological constant not to spoil relaxation we require 
\be
\label{M1bound}
M_1 \lesssim \Lambda_0^{1/4} \sim 10^{-3}~\text{eV} \; .
\ee
Since the scale $M_1$ determines the cutoff of the $\phi_1$ theory, it imposes an upper bound on the maximal value of the cosmological constant to be relaxed
\beq
\label{lambdastar}
\Lambda_{\star}\equiv 3\mpl^2H_{\star}^2~\lsim~\mpl^3 H_0 \sim (1\text{ TeV})^4~.
\eeq
We will assume that in the scenario with slow NEC violation the characteristic scales of the scanning sector and the NEC violating sector are similar, $M_1 \sim M_2 \sim \Lambda_0^{1/4}$.

In the second scenario the scanning of the cosmological constant continues. To remain within the regime of validity of the EFT we can assume that NEC violation kicks in when the deviation from $\dot\phi_1 = M_1^2$ is still moderately small 
\be
\( \frac{\dot \pi_1} {M_1^2}\)_{H=H_0} = x \lesssim 1 \;.
\ee
Notice, however, that $x$ cannot be very small since it is difficult to imagine how a tiny variation of $\dot\phi_1$ can induce a phase transition.
At this point, $\phi_1$ continues to scan the cosmological constant at a rate given by Eq.~\eqref{pi0}
\be
\epsilon_0 \sim x  \,\frac{M_1^4}{\Lambda_0}\; .
\ee
The NEC-violating accumulation of the energy density and reheating in this scenario thus has to be fast and complete within the time of order $\( \epsilon_0 H_0\)^{-1}$ --- before the effective cosmological constant further decreases by a significant amount. We provide an example of how this can happen in Sec.~\ref{sec:fastnec}. 
The bound on the maximal value of the relaxed cosmological constant~\eqref{lambdastar} in this case reads
\be
\Lambda_\star \lesssim 3 \mpl^2 M_1^2 \sim \mpl^2 \Lambda_0^{1/2} \(\frac{\epsilon_0}{x} \)^{1/2} \sim (1~\text{TeV})^4 \,  \(\frac{\epsilon_0}{x} \)^{1/2} \; .
\ee
Thus, in order to avoid significantly reducing $\Lambda_\star$, it is necessary to assume that the energy density of the scanning field evolves with a not-too-small $\epsilon_0$. On the other hand, the non-vanishing value of $\epsilon_0$ is related to the current dark energy equation of state parameter $(w_{DE} + 1)$, constrained to be less than $0.05$~\cite{Aubourg:2014yra}.

One could worry that in the second scenario the effective cosmological constant will continue to scan values much smaller than $\Lambda_0$. This however is not the case since $\phi_1$ remains on the slow-roll trajectory only for the time of order $\( \epsilon_0 H_0\)^{-1}$, after which $\epsilon$ becomes of order one and the system exits the slow-roll regime. Thus, the magnitude of the scanned effective cosmological constant never falls below its value at the moment of the breakdown of slow-roll (see Eq.~\eqref{gcslowroll})
\be
\label{furtherdown}
\Lambda_{\text{min}}\simeq \epsilon_0^{2/3} \Lambda_0 \; .
\ee
After the breakdown of slow-roll, the evolution of $\phi_1$ drives the potential energy to negative values,  causing the expansion of the universe to be followed by a fast contracting phase that ends in a collapse within a Hubble time at that moment, $H_\epsilon^{-1} \sim \epsilon_0^{-1/3} H_0^{-1}$.

Notice that the displacement of $\phi_1$ to relax $\Lambda_\star^4$ is of order $\Delta\phi \sim  \Lambda_\star^4/(H_0 M_1^2)$, using eq.~\eqref{lambdavalue}. This gives a disturbingly large displacement $\Delta\phi_1/\mpl \simeq \mpl/H_0$ using eqs~\eqref{M1bound} and \eqref{lambdastar}.

An important difference between the two scenarios is that in the former the Universe gets eventually stuck in a de Sitter vacuum: one has eternal inflation in the future. This reintroduces to some extent the measure problems. In the second scenario, on the other hand, the scanning continues and one eventually ends up in a AdS vacuum, so that eternal inflation is avoided. 

We stress that independently of the fate of $\phi_1$, our scenario can relax at most a vacuum energy of order $(\rm{TeV})^4$. Reducing higher values would require understanding the UV completion of the scanning field, or invoking other means of cancellation, \textit{e.g.}~supersymmetry, broken not too far from the  $\rm TeV$ scale or another relaxation mechanism.  Notice that the bounds on $\Lambda_\star$ discussed in this Section are much stronger than the ones based on stability we discussed above.

In this paper we do not attempt to study the phase transition to the NEC-violating regime. We wish however to argue that its details should not modify the global picture of the scenario. One could imagine the transition to proceed by nucleation of regions with the new phase or by a smooth cross-over to the NEC violating regime. In the first case the bubbles --- or whatever describes the nucleation of the new phase --- will give rise to a very inhomogeneous universe on scales much shorter than Hubble, while the long period of relaxation guarantees homogeneity on larger scales. Since the energy released in the phase transition is of the order of the cosmological constant, the total energy in the inhomogeneities cannot largely exceed the vacuum energy. We thus argue these inhomogeneities will be quickly erased as the universe is accelerating and soon it will even violate the NEC (the problem is similar to the start of inflation in the presence of initial inhomogeneities; for a recent study see \cite{East:2015ggf}). The other option is that the phase transition is smooth like the waterfall transition at the end of hybrid inflation. In this case we do not expect the formation of large inhomogeneities even on short scales.

\section{Slow violation of the NEC} \label{sec:slownec}
\begin{figure}[h!]
\center
\includegraphics[width=12cm]{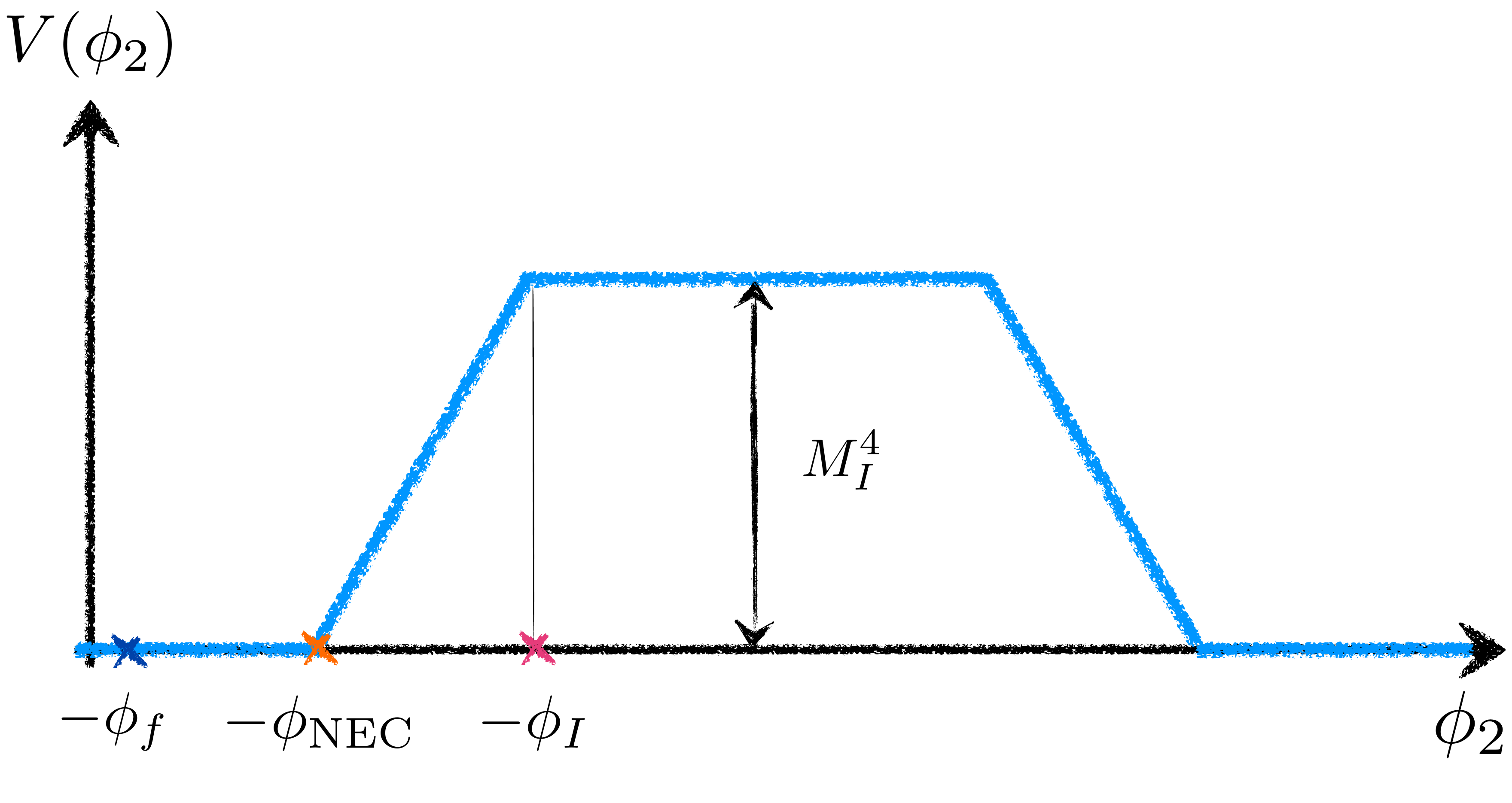}%
\caption{The potential $V(\phi_2)$ for the NEC-violating scalar. The figure is not to scale: all slopes are extremely small compared to the scale setting the potential's height.}
\label{fig:vphi}
\end{figure}

In the case that the evolution of the scanning field $\phi_1$ stops as a result of the phase transition, $\phi_2$ has essentially unlimited amount of time for building up the inflationary energy density through its NEC-violating dynamics. The Lagrangian for the latter field that we will implement for these purposes is then similar to \eqref{ghost}
\beq
\label{ghost2}
S_{\phi_2} = \int d^4 x \sqrt{-g} \bigg[M_2^4 P_2\(X_2\)-V(\phi_2)  +\dots\bigg ],
\eeq
where the shape of the potential $V(\phi_2)$ is sketched in Fig.~\ref{fig:vphi}. After acquiring a non-zero velocity ($\dot \phi_2 = M_2^2$), $\phi_2$ starts rolling towards a piecewise linear potential. We assume that the theory (including the potential) enjoys a global $Z_2$ symmetry under reflection of the field with respect to the origin, $\phi_2\to -\phi_2$. Moreover, we will assume that \textit{in the shift symmetric region}, the theory is also invariant under reflections of $\phi$ around a generic point in the field space.\footnote{Notice, that this property is not implied by the global $Z_2$ symmetry. For example, the operator $\tanh(\phi_2) (\partial\phi_2)^2\Box\phi_2$ is allowed by the global $Z_2$ symmetry, but reduces to an operator with an odd number of fields in the shift-symmetric region.} In this way the theory is invariant under the flip of sign of the field velocity in the shift symmetric region. The combination of these two symmetries implies that the system has the same vacuum energy before and after the feature in the potential  (see Fig.~\ref{fig:vphi}).\footnote{In the next Section we will employ a different symmetry for this purpose.} 
The presence of a feature in the potential for a certain range of $\phi_2$ will not spoil the shift symmetry away from it, since renormalization is local in field space. 

Upon climbing up the positive slope, $\phi_2$ slowly builds up energy, thereby violating the NEC \cite{Creminelli:2006xe}. We will denote the maximal energy density created this way by $M^4_I$. The validity of the EFT of $\phi_2$ bounds $M_I$: $M_I^2/\mpl \lesssim M_2$ that gives $M_I \lesssim$ TeV using Eq.~\eqref{m2_max}. After NEC violation has ceased and $\phi_2$ finds itself on the plateau on top of the potential, its coupling to some ordinary matter field $\chi$ is assumed to activate, by means of which the latter acquires potential energy. At this point the universe is assumed to undergo inflation and (after $\phi_2$ rolls back to the region with a vanishing potential) to reheat through the dynamics of $\chi$. This connects our scenario to the standard Big Bang cosmology, which, at late times (and after all possible phase transitions have happened) finds itself in a state with a small cosmological constant $\Lambda_0$. 
We will postpone discussing how inflation and reheating fit into this picture until Sec.~\ref{sec:reheating}.

On the way up the positive slope, one has to deal with the known problems associated with NEC violation (see, \textit{e.g.} \cite{Hsu:2004vr,Creminelli:2006xe}). In the particular case of the ghost condensate there are two possible issues: besides the Jeans-like instability discussed above Eq.~\eqref{jeans} there is a gradient instability associated with NEC violation. The rates corresponding to these are respectively $
\omega_{\text {Jeans}} \sim M_2^3/\mpl^2$
and $\omega_{\rm grad} \sim \dot H \mpl^2/M_2^3$~.
These instabilities are harmless if these rates are less than the expansion rate of the universe, which imposes the following bounds on $M_2$ \cite{Creminelli:2006xe}
\beq
\label{bounds}
\frac{\dot H}{H}~\lsim~ \frac{M_2^3}{\mpl^2}~\lsim~ H~.
\eeq
We have encountered an analogous upper bound in the context of the scanning field $\phi_1$; it yields a constraint, similar to  Eq. \eqref{jeans}, $
M_2~ \lsim ~(\mpl^2 H_0)^{1/3}\sim 10 ~\text{MeV}$.
This is a much weaker constraint than the one we derived in Eq. \eqref{m2_max} based on naturalness of the NEC-violating phase transition. 
The lower bound on $M_2$ that follows from \eqref{bounds}, on the other hand, strongly constrains the slope of the linear piece of the potential, $V'(\phi_2)=\lambda_2^3$,
\beq
\label{lambda2upper}
\lambda_2^3 ~\lsim~ M_2 H_0^2~.
\eeq
Note that the above constraint forces $\lambda_2$ to be extremely small compared to the height of the potential $M_I$. Indeed, assuming that the latter scale takes on its maximal value, $M_I^4\sim \Lambda_\star$, we have
\be
\label{lambda2upper2}
\frac{\lambda_2}{M_I}~\lsim~\frac{(M_2H_0^2)^{1/3}}{(\mpl^3H_0)^{1/4}}\sim\left(\frac{\Lambda_0^{1/4}}{\mpl}\right)^{7/6}\sim 10^{-35}\;.
\ee
As in the case of the scanning field $\phi_1$, the smallness of $\lambda_2$ is technically natural given that it is a spurion of $\phi_2$ shift symmetry breaking.

The smallness of $\lambda_2$ significantly limits the speed at which energy density can be built up by the dynamics of the NEC-violating sector. One can readily estimate how long it takes for $\phi_2$ to roll up the linear slope. Assuming that the energy density grows all the way up to $M_I^4\sim\Lambda_{\star}$ in the process, we have
\beq
\rho\simeq \int \dot\rho ~dt \sim \lambda_2^3 M_2^2 t\sim \Lambda_{\star}~.
\eeq
Making use of Eqs. \eqref{m2_max}, \eqref{lambdastar} and \eqref{lambda2upper}, the resulting time scale can be expressed as
\beq\label{time}
t~ \gtrsim ~\frac{1}{H_0}~ \frac{\Lambda_{\star}}{M_2^3 H_0} \sim \frac{10^{90}}{H_0}~.
\eeq
The time required to create a sizeable amount of energy density is way beyond the Hubble time right before the onset of the NEC-violating phase transition. This is due to the smallness of the cutoff $M_2$ compared with the energy density we want to produce. Slow violation of the NEC is therefore only relevant if the scanning of the cosmological constant stops completely as a result of this transition. Much faster violation of the NEC is possible, provided the cutoff evolves in time, as we will study in the next Section.

\section{Fast violation of the NEC}\label{sec:fastnec}

Rather than stopping, the scanning field may retain a non-zero speed $\dot\phi_1\sim M_1^2$ after the transition to the NEC-violating phase. In this case it is crucial that this phase followed by inflation and reheating complete relatively fast, within a time of order, or less than $(\epsilon_0H_0)^{-1}$. This will guarantee that, by the time the universe reaches its present state, the cosmological constant has not been reduced by a significant amount. 

\begin{figure}[t!]
	\center
	\includegraphics[width=15cm]{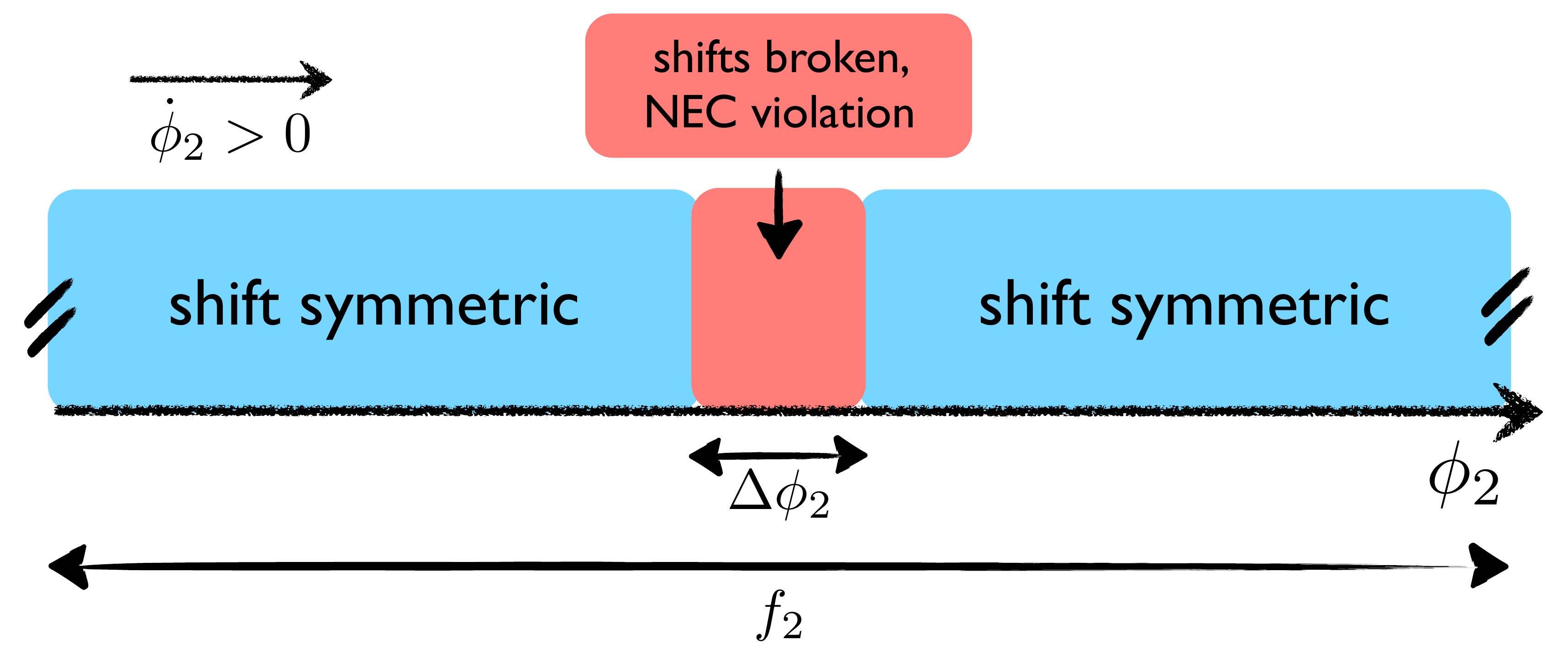}%
	\caption{The structure of the $\phi_2$ effective theory (the endpoints of the depicted field space are identified). }
	\label{fig:phi2}
\end{figure}

For the purposes of achieving fast NEC violation, we will again rely on a theory that in most of the $\phi_2$-field space is described by a ghost condensate-like shift-symmetric action.
The shift symmetry is only broken in a narrow interval of width $\Delta \phi_2$, centered at $\phi_2=0$, see Fig. \ref{fig:phi2} for an illustration. This region is also where the violation of the NEC happens in our model, as we discuss shortly.
A successful implementation of the scenario requires that the $\phi_2$ field space be periodic with a period $f_2$, so that any $\phi_2$ is identified with $\phi_2+f_2$. 
The virtue of periodicity is twofold.
First, periodicity, together with invariance under internal shifts, imposes that away from the red region with broken symmetry, the value of the cosmological constant is identical on the two sides of that region,
see Fig.~\ref{fig:phi2}. Second, periodicity of $\phi_2$ is important for naturalness of the model under consideration. This is because, given the limited time for the NEC-violating phase to complete, $\phi_2$ has to hit the red region with broken symmetry within a time that does not parametrically exceed $(\epsilon_0H_0)^{-1}$ (we are assuming the phase transition will leave $\phi_2$ in a generic point of the field space).\footnote{Given the periodicity, one can imagine to have a cyclic evolution of the universe  \cite{Steinhardt:2006bf} which goes through NEC violation many times. As discussed around eq.~\eqref{furtherdown}, the cosmological constant will continue the scanning only until the breaking of slow-roll, so that its value will not be parametrically different in the various cycles, unless $\epsilon_0$ is not very suppressed.} Since $\phi_2$ moves with a constant speed $\dot\phi_2\sim M_2^2$ in the shift-symmetric region, this translates, assuming $M^4_2\sim\mpl^2H_0^2 $ (see Eq.~\eqref{m2_max}), and $\epsilon_0 \sim 1$, into $f_2 \lesssim \mpl$. This is reminiscent of the constraint on axion decay constants that follows from the weak gravity conjecture of Ref.~\cite{ArkaniHamed:2006dz}. 

\subsection{A method for constructing strongly NEC-violating cosmologies}

To complete the picture we have to come up with an explicit example of a theory with the symmetry structure illustrated in Fig.~\ref{fig:phi2} that would lead to a strong violation of the NEC. Rather than constructing solutions to a particular theory that conforms to the asymptotic symmetries of interest, we will employ a trick whereby the appropriate theory itself is reverse-engineered based on a postulated ansatz for the desired cosmological evolution. For our purposes, this will provide a mechanism to smoothly build up the inflationary energy density within a period $t_{\cancel{NEC}}~\lsim ~H_0^{-1}$ --- at the same time avoiding instabilities and superluminal perturbations usually associated with NEC violation. Our presentation draws heavily on an analogous construction of Ref. \cite{Pirtskhalava:2014esa}.
 
Inspired by the fact that NEC violation is possible in Galileon theories, let us consider a theory of the following form (it is convenient to work with a dimensionless angle, defined as  $\theta \equiv \phi_2/f_2$):
\be
\label{ggg}
S_\theta=\int d^4 x ~\sqrt{-g} ~\bigg[f_2^2 \mathcal{F}_1(\theta) (\p\theta)^2+\frac{f_2^3}{M_\theta^3}\mathcal{F}(\theta)(\p\theta)^2\Box\theta +\frac{f_2^3}{2M_\theta^3} \mathcal{F}_2(\theta) (\p\theta)^4-V(\theta)    \bigg ]~.
\ee 
To reheat the universe, we are going to need an extra scalar, $\chi$, which will act as a waterfall field; we postpone this discussion to the following Section and consider only the field $\theta$ here. $\mathcal{F}_{1,2}$ and $\mathcal{F}$ are \textit{a priori} arbitrary dimensionless functions of $\theta$. The typical values of the decay constant $f_2$ we will have in mind are around the Planck scale, and often we will simply assume $f_2 = \mpl$. Since the coefficients of the operators are arbitrary we can choose for later convenience
\be
\label{lambdac}
M_\theta^3  = \frac{3}{2}f_2H_0^2~.
\ee

Furthermore, we will be interested in the functions $\mathcal{F}_{1,2}$ and $\mathcal{F}$ such that the resulting theory complies with the symmetry requirements illustrated in Fig.~\ref{fig:phi2}. In particular these functions must be constant asymptotically with the same value on the two sides of the shift-breaking region. For reasons that will become clear below, we will also assume that $\mathcal{F}$ is very small in the shift-symmetric part of the field space (while we will have $\mathcal{F}_1 \sim \mathcal{F}_2 \sim 1$). What this means in practice is that the higher-derivative Galileon has the right magnitude dictated by the derivative expansion, and it thus gives sub-leading effects compared to the more relevant one-derivative operators.\footnote{In particular, canonical normalization of the action \eqref{ggg} for $\mathcal{F}_1\sim\mathcal{F}_2\sim 1 $ reveals that the cubic operator is suppressed by powers of the scale $(f_2 H_0^2/\mathcal{F})^{1/3}$, while the quartic one --- by powers of $(f^2_2 H_0^2)^{1/4}$. Requiring the two scales to be comparable as dictated by na\'ive dimensional analysis (and recalling that $f_2\sim \mpl\gg H_0$) yields an extremely suppressed value of $\mathcal{F}$ in the shift-symmetric region.}
We will have to break this na\'ive power counting in the NEC-violating region with broken shifts, where the  Galileon operator will play a crucial role. Notice it is consistent and technically natural to have a large $\Box \theta (\partial\theta)^2$ operator compared to the $(\partial\theta)^4$ since the first one is Galilean invariant, while the second is not \cite{Pirtskhalava:2015nla}. Regarding the potential, we are going to look for solutions such that the potential after the shift-breaking region is much larger than before: the idea is that the waterfall field $\chi$ gets trapped in some minimum with higher potential energy during the NEC-violating phase. It will eventually return to the same vacuum as before (with a very small, relaxed cosmological constant), releasing the energy into the thermal bath. Therefore, although the potential for $\theta$ is periodic, the potential energy will be higher until $\chi$ drops to the true minimum.
We will come back to this part of the model in Section \ref{sec:reheating}.

The dynamics of the system \eqref{ggg} is governed by the Einstein's equations plus the scalar equation of motion. These however are not independent: as a consequence of diffeomorphism invariance, the scalar equation can be traded for the conservation of its stress-energy tensor via
\beq
\label{s.eom}
\nabla_\mu T^\mu_{~\nu}=-\frac{1}{\sqrt{-g}}\frac{\delta S_\theta}{\delta \theta} \p_\nu\theta~.
\eeq
On homogeneous FRW backgrounds, it is the energy conservation, $\dot \rho + 3H (\rho+p)=0$ that yields the $\theta$-equation of motion. Energy conservation on the other hand follows from the temporal and space components of the Einstein's equations. Therefore, one can choose the latter two to make up the complete system determining the background evolution. 

The expressions for the energy density and pressure due to a homogeneous evolution $\theta(t)$  are
\ba
\label{rho}
\rho &=&\frac{f_2^2}{H_0^2} ~\dot \theta^2 \big [ \mathcal{F}_2(\theta)\dot\theta^2+4  \mathcal{F}(\theta) H\dot\theta -  H_0^2\mathcal{F}_1(\theta)\big ]-\frac{2}{3}\frac{f_2^2}{H_0^2}\mathcal{F}'(\theta)\dot\theta^4 +V(\theta)~,\\
\label{press}
p &=& \frac{f_2^2}{3 H_0^2}~\dot\theta^2 \bigg[ \mathcal{F}_2(\theta) \dot\theta^2-4\mathcal{F}(\theta)\ddot \theta-3H_0^2\mathcal{F}_1(\theta) \bigg]-\frac{2}{3}\frac{f_2^2}{H_0^2}\mathcal{F}'(\theta)\dot\theta^4-V(\theta)~,
\ea
where $'$ denotes differentiation with respect to the argument (so that $\dot{\mathcal{F}}=\mathcal{F}'\dot\theta$).
The two functions $\mathcal{F}_{1,2}(\theta)$ can be solved for with the help of the Friedmann equations, $3 \mpl^2 H^2=\rho$ and $\mpl^2 (3 H^2+2 \dot H)=-p$, which yield
\beq
\label{f1}
\mathcal{F}_1&=&\frac{18 \mpl^2 H_0^2 H^2+9\mpl^2 H_0^2\dot H-6 f_2^2 \mathcal{F} H\dot\theta^3-6 f_2^2  \mathcal{F} \dot\theta^2\ddot\theta-2f_2^2\dot{\mathcal{F}}\dot\theta^3-6 H_0^2 V}{3 f_2^2 H_0^2\dot\theta^2}~,\\
\label{f2}
\mathcal{F}_2&=&\frac{9\mpl^2 H_0^2 H^2+3\mpl^2 H_0^2\dot H-6f_2^2\mathcal{F} H\dot \theta^3-2f_2^2 \mathcal{F} \dot\theta^2\ddot\theta-3 H_0^2 V}{f_2^2 \dot\theta^4}~.
\eeq  
Now, for any \textit{postulated} homogeneous profile for  $\theta$, $\mathcal{F}$, the Hubble rate $H$, and the scalar potential $V$, one can find a theory (\textit{i.e.} find $\mathcal{F}_{1,2}(\theta)$) such that the chosen background solves its equations of motion. The recipe for constructing the relevant solutions goes as follows: i) pick arbitrary background profiles for $\theta(t)$, $\mathcal{F}(t)$, $H(t)$, and the potential $V(\theta(t))$, ii) for the chosen profiles, find the time-dependent functions $\mathcal{F}_{1,2} (t)$  with the help of \eqref{f1} and \eqref{f2}, iii) invert the expression for $\theta(t)$ to find $t=t(\theta)$ (we do not have problems of inversion if we remain within a single period of $\theta$), and iv) using the previous steps find $\mathcal{F}_{1,2}$ as functions of the dynamical field, rather than time: $\mathcal{F}_{1,2}=\mathcal{F}_{1,2}\(t(\theta)\)$.
Importantly, we should check whether a given cosmological solution obtained through the above procedure is stable and devoid of superluminal perturbations. A detailed analysis of perturbations for the theory \eqref{ggg} can be found in Ref. \cite{Pirtskhalava:2014esa}, and here we will only present the relevant expressions without deriving them. 
\subsection{Perturbations and stability}
In the unitary gauge defined by the absence of $\theta$ - fluctuations, the system's only scalar degree of freedom is captured by the curvature perturbation on uniform-density hypersurfaces
\beq
\label{gij}
g_{ij}=a(t)^2 (1+2\zeta)\delta_{ij}~.
\eeq 
A peculiar feature of the theory \eqref{ggg} --- in particular of the Galileon operator --- is that it leads to second-order equations of motion both for the scalar and for metric on an arbitrary background. Related to that, the quadratic $\zeta$ action takes on the standard two-derivative form
\be
\label{quadact}
S_\zeta=\int d^4x~ a^3~\bigg[A(t)~\dot\zeta^2-B(t)~\frac{1}{a^2}\(\p\zeta\)^2   \bigg]~.
\ee
The kinetic coefficients $A$ and $B$ are given by \cite{Creminelli:2006xe,Pirtskhalava:2014esa} 
\begin{align}
\label{A}
A(t) &=\frac{\mpl^2 (-4 \mpl^4 \dot H-12\mpl^2 H \hat M^3+3\hat M^6+2\mpl^2 M^4)}{(2\mpl^2 H-\hat M^3)^2}~,\\
B(t)&=\frac{\mpl^2 \(-4 \mpl^4 \dot H+2\mpl^2 H \hat M^3-\hat M^6+2\mpl^2\p_t\hat M^3\)}{(2\mpl^2 H-\hat M^3)^2}~,
\end{align}
where $M^4$ and $\hat M^3$ have been defined as follows 
\beq
\label{M's}
M^4(t)=\frac{4}{3} \frac{f_2^2}{H_0^2} \(2 \mathcal{F}_2(\theta) \dot\theta^4+\mathcal{F}\dot \theta^2\ddot\theta+9\mathcal{F} H\dot\theta^3\)-\frac{4}{3}\frac{f_2^2}{H_0^2}\dot{\mathcal{F}}\dot\theta^3, \quad \hat M_3^3(t)=\frac{4}{3}\frac{f_2^2}{H_0^2}\mathcal{F}\dot\theta^3~.
\eeq
With the above expressions at hand, we are in a position to postulate an arbitrary NEC-violating set of profiles $H(t)$, $\mathcal{F}\(\theta(t)\)$, $\theta(t)$ and $V\(\theta(t)\)$ and check the properties of perturbations on the chosen background. In particular, the absence of ghost and gradient instabilities require $A>0$ and $B>0$, while subluminal propagation of $\zeta$ imposes $B/A\leq 1$. 

\subsection{An ansatz}
An ansatz that we will find particularly useful for fast NEC violation is defined by the following set of relations
\beq
\label{ansatz}
\dot\theta(t) = H_0~,\quad \mathcal{F}(t)=\alpha(t)\frac{H(t)}{H_0}, \quad  \dot H(t) =\varepsilon(t) H(t)^2~,\quad V(t)=3\kappa(t) \mpl^2 H(t)^2~,
\eeq
where $\alpha(t)$, $\varepsilon(t)$, $H(t)$ and $\kappa$(t) are yet unspecified functions.\footnote{Note the unconventional sign in the definition of $\varepsilon$ in \eqref{ansatz} as opposed to the slow-roll parameter, $\epsilon=-\varepsilon$, we worked with in the previous sections.} The function $\alpha(t)$ quantifies the role of the Galileon operator in \eqref{ggg}, while $\kappa(t)$ determines the fraction of the total energy density $\rho$, stored in the potential, $\kappa=V/\rho$. 
A rather attractive feature of this ansatz is that the kinetic coefficients $A$ and $B$ take a simple form on it
\ba
\label{akin}
A &=& 3\mpl^2~ \frac{36+4\alpha^2-2\alpha (\varepsilon+9)-2\beta-36\kappa+9\varepsilon}{(3-2\alpha)^2}~, \\
\label{bkin}
B&=& \mpl^2 ~\frac{6\alpha(1+\varepsilon)+6\beta-4\alpha^2-9\varepsilon}{(3-2\alpha)^2}~,
\ea
where we have set $f_2=\mpl$ for simplicity, and have defined 
\beq
\label{beta}
\beta \equiv \frac{1}{H}\frac{d}{dt} \alpha~.
\eeq

Note that according to \eqref{ansatz} the scalar never changes its velocity after the phase transition, always moving with $\dot\theta = H_0$. For field values far from the shift symmetry-breaking region in Fig.~\ref{fig:phi2}, if we assume $\alpha = 0$, this describes a ghost condensate solution.
Prior to hitting that region, there is no potential energy ($\kappa=0$), while the kinetic energy of $\theta$ is
\be
\label{thetakin}
\rho_\theta = 3\mpl^2 H_0^2~,
\ee 
consistently with having a de Sitter spacetime with the Hubble rate $H_0$ (for simplicity of presentation, we ignore all other sources of energy except for the sector responsible for reheating, see below).
Upon approaching the region with broken shifts from the left, NEC violation gradually turns on ($\varepsilon(t)>0$), and the Hubble rate starts increasing. Subsequently, when the NEC-violating dynamics ceases and $\theta$ rolls back into the shift-symmetric part of the field space, the curvature of the universe becomes constant again. However, it is not the same as prior to NEC violation. Incorporating reheating (see Sec.~\ref{sec:reheating}) requires that the Hubble rate after NEC violation, $H_I$, be much larger than $H_0$. Moreover, the energy density at this stage is almost fully due to the \textit{potential energy} of the reheating sector, meaning that $\kappa \simeq 1$ to a very good approximation. 
Eventually $\theta$ should return to the same state after the NEC-violating phase (recall that the Lagrangian of $\theta$ is periodic), where it still contributes to the energy density of the universe by the amount $\rho_\theta$ in \eqref{thetakin}.\footnote{This is where the absence of the Galileon operator outside the shift symmetry-breaking region is important --- see discussion below for more detail.} 
The Friedmann equation right after the NEC-violating phase can then be written as 
\be
\label{postnecfied}
3\mpl^2 H_I^2 =3\kappa \mpl^2 H_I^2+3\mpl^2H_0^2~.
\ee

As an instructive sanity check of our procedure, let us verify that $\theta$ indeed returns to the same vacuum after it exits the region with broken shifts. As we have chosen the field velocity not to change at all for our ansatz \eqref{ansatz}, this reduces to verifying that the effective field theory of $\theta$ itself (that is, $\mathcal{F}_1$ and $\mathcal{F}_2$) is the same before and after NEC violation. The latter fact is not so manifest, since $\mathcal{F}_1$ and $\mathcal{F}_2$ apparently depend on the Hubble rate in \eqref{f1} and \eqref{f2}, which \textit{does} change as a result of the NEC-violating phase as we have just discussed. Prior to the latter phase Eqs. \eqref{f1} and \eqref{f2}, upon setting $H=H_0$ and $\alpha=\beta=\varepsilon=\kappa=0$, imply 
\be
\label{beforenec}
\mathcal{F}^{\rm before}_1=6, \qquad \mathcal{F}^{\rm before}_2=9~.
\ee
\textit{After} NEC violation, on the other hand, plugging  in the same values of the parameters except that now $H=H_I$ and $\kappa\neq 0$, yields
\be
\label{afternec}
\mathcal{F}^{\rm after}_1= 6 (1-\kappa) \frac{H_I^2}{H_0^2}, \qquad \mathcal{F}^{\rm after}_2= 9 (1-\kappa) \frac{H_I^2}{H_0^2}~.
\ee
It suffices to use Eq. \eqref{postnecfied} to show that \eqref{beforenec} and \eqref{afternec} imply $\mathcal{F}^{\rm before}_1=\mathcal{F}^{\rm after}_1$ and $\mathcal{F}^{\rm before}_2=\mathcal{F}^{\rm after}_2$. 
Note, however, that the same logic would not go through had we allowed for the Galileon (non-zero $\mathcal{F}$) in the shift-symmetric region of field space. This is because the energy density due to this operator depends on the Hubble rate, see Eq. \eqref{rho}. Therefore, the drastically different value of $H$ after NEC violation would constitute an obstruction --- at least for the simple ansatz \eqref{ansatz} --- to returning to the same EFT on the right of the shift symmetry-breaking region.

\subsection{How to strongly violate the NEC within a stable and subluminal effective field theory}

In principle, we can choose an arbitrary profile for the Hubble rate, such that it takes on the small value $H_0$ before the field hits the NEC-violating region of the field space, while growing abruptly --- all the way up to $H_I$ --- within the latter region. The time required to produce a significant amount of energy density this way can be roughly estimated as
\be
t_{\cancel{NEC}}\sim \frac{1}{\varepsilon_{\cancel{NEC}} H_0}
\ee
where $\varepsilon_{\cancel{NEC}}$ is a typical value of the slow-roll parameter $\varepsilon(t)$ over the NEC-violating period $t_{\cancel{NEC}}$. 

\begin{figure}[t!]
	\center
	\includegraphics[width=14cm]{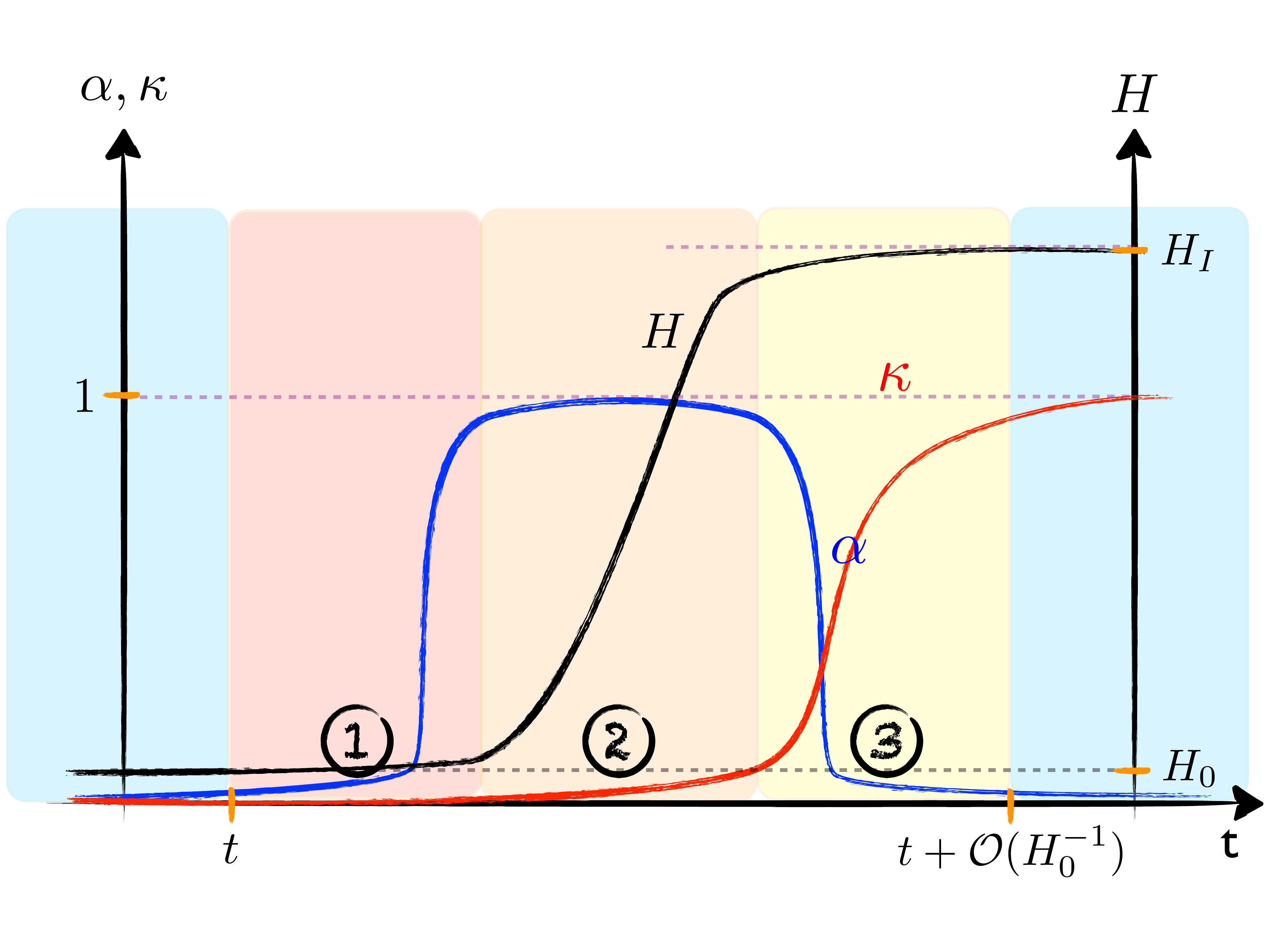}%
	\caption{A sketch of the time dependence of various functions characterizing the ansatz \eqref{ansatz} in the three-stage NEC-violating phase.}
	\label{fig:ansatz}
\end{figure}

A non-trivial task is to find a solution of this type, such that it does not lead to any sort of instability or superluminal perturbations all along the cosmological evolution. Fast NEC violation can only be achieved in a stable way in the presence of the Galileon operator, and we will thus need the parameters $\alpha$ and $\beta$ to turn on in the NEC-violating region.
Furthermore, as we have discussed above, both have to decay outside this region of field space. Given that its derivative, $\beta=H^{-1}\dot\alpha$, enters into the dynamics, the decay of $\alpha$ should be smooth. On the other hand, at least formally nothing requires $\beta$ to be continuous, and we will make it a piecewise function for the sake of presenting examples of stable and subluminal cosmologies with fast NEC violation. It is clear that these examples can always be deformed into versions with a smooth $\beta$ without compromising stability and/or subluminality. 

Our scenario for fast, stable and subluminal NEC violation is described in Fig.~\ref{fig:ansatz} and goes as follows.

\begin{itemize}

\item
Stage 1: By continuity, after $\theta$ hits the region with broken shifts, $\alpha$ is very small for a while. We will also assume that $\varepsilon$ is small, so that the kinetic coefficients in \eqref{akin} and \eqref{bkin} read
\be
A=\frac{2}{3}\mpl^2 \big(18(1-\kappa)-\beta\big), \qquad B = \frac{2}{3}\mpl^2\beta~.
\ee
Even for $\kappa =0$, there is significant parameter space, consistent with stability and subluminality of scalar perturbations:
\be
0 <\beta \leq 9 \qquad \text{(stability and subluminality)}~.
\ee
This means, using Eq. \eqref{beta}, that the parameter $\alpha$ can be smoothly increased to an order-one value within a time of order $H^{-1}_0$.
\item
Stage 2: When $\alpha$ becomes order-one, $\beta$ turns off for a while. For $\beta=0$, the parameter space consistent with i) positivity of $A$ and $B$ and  (sub)luminality of $\zeta$'s speed of propagation, $c_s^2=B/A$, ii) positivity of the potential energy ($\kappa >0$) and iii) NEC violation ($\varepsilon>0$) can readily be found. We will not fully reproduce it here, merely quoting its part that is continuously connected with the Stage 1:
\begin{align}
\begin{split}
\label{ps1}
0<\alpha<\frac{3}{2}~\bigwedge~0<\varepsilon<\frac{2}{3}\alpha ~\bigwedge~ 0<\kappa\leq \frac{1}{27}\(27+4\alpha^2-3(\varepsilon+5)\alpha+9\varepsilon\)~.
\end{split}
\end{align}
It is clear from \eqref{ps1} that fully stable and subluminal NEC-violating solutions (with a speed of sound of order unity all along) exist for an order-one $\varepsilon$ --- meaning that one can build up (arbitrarily large) inflationary energy density $\rho_I \simeq 3\mpl^2 H_I^2$ within a time of order $H_0^{-1}$.
Note, that according to Eq. \eqref{ps1}, an order-one $\varepsilon$ is only possible for an order-one $\alpha$; this is why we had to increase the latter parameter before a significant rate of NEC violation were allowed.
\item
Stage 3: Having built up the inflationary energy density, $\theta$ has to return to its original state, but now in a universe with the large curvature $H_I$ driven by the \textit{potential energy} of the sector responsible for reheating. To this end, we have to make sure that $\alpha$, $\beta$ and $\varepsilon$ all drop back to zero, while $\kappa$ asymptotes to one, as discussed around Eq. \eqref{postnecfied} and illustrated in Fig. \ref{fig:phi2}. That $\alpha$ decreases, means that $\beta$ has to turn negative at this stage. Imposing again all of the requirements that led to Eq. \eqref{ps1} --- but now with a non-zero and negative $\beta$ --- yields the allowed parameter space. Again, this parameter space is quite large and not particularly illuminating, so we will not reproduce it here. For our purposes, if suffices to note that it includes as a subset Eq. \eqref{ps1}, supplemented by
\be
\label{betaspace}
\frac{1}{6}\(9\varepsilon-6\alpha+4\alpha^2-6\alpha\varepsilon\)<\beta<0~.
\ee
With the latter formula at hand, one can evolve the parameter $\alpha$ from its order-one value back to zero --- with a time derivative that satisfies \eqref{betaspace} all along. Moreover,  $\varepsilon$ can  also be dialed to zero in the process, in a way that is compatible with the condition $0<\varepsilon<2\alpha/3$ at any given moment of time. One can thus easily send the NEC-violating sector back to its original state in a fully stable/subluminal manner --- while keeping the Hubble rate $H_I$ much larger than what it was prior to NEC violation. 
The last thing to check is that, given the limitedness of the rate at which $\alpha$ can decrease, Eq. \eqref{betaspace}, this parameter can be dialed to zero within a sufficiently short time. Let us for simplicity of the argument assume, that $\varepsilon$ and $\alpha$ are of the same order and that they start evolving towards zero from an initial value, somewhat smaller than one. In that case, one can keep only the linear terms on the left hand side of Eq. \eqref{betaspace}, and (almost) saturating this inequality yields $\beta \simeq -c\alpha$, where $c$ is some order-one number. Recalling the definition \eqref{beta}, we have
\be
\alpha = \alpha_i e^{-cH_I t}~,
\ee
meaning that $\alpha$ can be driven to zero within a time of order $H^{-1}_I\ll H_0^{-1}$. 
\end{itemize}

It is unlikely that Eq.~\eqref{ansatz} is the unique ansatz, consistent with our stringent constraints on the dynamics of the theory in the NEC-violating phase, and other solutions with even faster NEC violation may well be found.
For our purposes, however, the ansatz \eqref{ansatz} perfectly does the job: it shows that the dynamics of $\theta$ does allow to build up a significant amount of inflationary energy density within a Hubble time $H_0^{-1}$ in a fully stable and subluminal way. Moreover, it does so in a way that naturally accounts for inflation and the subsequent reheating of the universe. We will return to discussing the latter aspect of our model in some more detail in Sec. \ref{sec:reheating}.

\subsection{The sliding cutoff}
Given that the Hubble rate increases drastically in the NEC-violating phase, one may wonder about the fate of the effective field theory description of $\theta$'s dynamics. Outside the NEC-violating region of the field space the low-energy EFT of $\phi_2$ is characterized by a very low cutoff bound by Eq.~\eqref{m2_max}. As can be inferred by canonically normalizing the field in \eqref{ggg}, $M_2\sim (f^2_2 H^2_0)^{1/4}$ (recall that $\mathcal{F}$ vanishes in that region, while $\mathcal{F}_1$ and $\mathcal{F}_2$ are field-independent order-one constants). The largest possible curvature that the universe experiences after relaxation is $H_I \lesssim H_\star \lesssim M_2$, meaning that the dynamics is well within the regime of validity of the EFT in the region under consideration. 
\textit{During} the NEC-violating phase, however, the Galileon operator turns on, and a strong violation of the NEC requires it to have a large Wilson coefficient. For instance, for order-one $\mathcal{F}$, $\mathcal{F}_1$ and $\mathcal{F}_2$, this operator would be suppressed by powers of the scale $M_\theta\sim (f_2 H^2_0)^{1/3}$ --- much lower than $H_\star$. This however does not happen in our scenario: $\mathcal{F}_1$ and $\mathcal{F}_2$ are much larger than unity in the bulk of the NEC-violating region, where $\mathcal{F}$ becomes sizeable.
Indeed, while $\mathcal{F}$ is proportional, for $\alpha\sim 1$, to the instantaneous Hubble rate $H(t)$ according to our ansatz \eqref{ansatz}, it follows from Eqs. \eqref{f1} and \eqref{f2} that $\mathcal{F}_1$ and $\mathcal{F}_2$ in fact grow faster
\be
\mathcal{F}_{1}\sim \frac{H(t)^2}{H_0^2}, \qquad \mathcal{F}_{2} \sim \frac{H(t)^2}{H_0^2}~. 
\ee
Furthermore, the time evolution of these functions is precisely such that upon canonical normalization of the field perturbation, $\theta_c \equiv f_2\sqrt{|\mathcal{F}_1|} \delta\theta$, the scale suppressing the cubic operator $(\p\theta_c)^2\Box\theta_c$ at any given moment is
\be
M_\theta(t)^3 \sim f_2 H(t)^2~,
\ee
instead of $M_\theta^3 \sim f_2 H_0^2$. That is to say, at any particular moment of time, the sliding cutoff of the theory $M_\theta(t)$ is much higher than the instantaneous value of the Hubble rate. Likewise, one can check that it is the sliding scale $(f_2 M_\theta(t)^3)^{1/4}\gg M_\theta(t)$ that suppresses the quartic interaction $(\p\theta_c)^4$ at any given time. 

This concludes our discussion of the mechanisms that allow to create the inflationary universe within the context of the relaxed cosmological constant. Before turning to the next Section to discuss the implementation of inflation and reheating in our setup a comment is in order. One may wonder whether all of the above discussion could be simplified by just focusing on the effective theory of perturbations around a given background, similarly to what done in \cite{Creminelli:2006xe}: indeed it would be straightforward to choose the operators which describe the quadratic perturbations in such a way as to avoid instabilities and superluminality. However in this approach it would be hard (impossible?) to impose that the action after the NEC violating phase be the same as before up to a shift in the potential energy: the extra potential energy changes the solution and this approach is not suited to compare different background solutions.

\section{\label{sec:reheating}Reheating the universe after NEC violation}
To complete the picture, one has to come up with a mechanism for transferring the energy density stored in the NEC-violating scalar $\phi_2$ into a sector that will eventually be responsible for reheating the universe and connecting it to the conventional Big Bang cosmology. To this end, we will invoke another canonical scalar $\chi$ coupled to $\phi_2$: the idea is that the energy created during NEC violation is stored in $\chi$ who gets stuck into a minimum with large energy. Eventually $\chi$ comes back to the true vacuum (the one with relaxed cosmological constant) reheating the universe in the same way as the waterfall field does after hybrid inflation.
 For definiteness, we will focus on the scenario with slow NEC violation where the ghost condensate slowly climbs up the piecewise linear potential $V(\phi_2)$ shown in Fig.~\ref{fig:vphi}, building up a potential energy density of order $M_I^4$. A similar reheating mechanism can be applied to the scenario with fast NEC violation.

Consider a canonical scalar field $\chi$ whose potential has the following form
\ba\label{potentialU}
U(\phi_2,\chi)=W(\chi)+g e^{-\chi/M_I} \cdot V(\phi_2)+\hat M^4 \cdot f\(\frac{\chi-\chi_I}{\hat M}, \frac{\phi_2}{\tilde M}\)~,
\ea
where $g$ is a small dimensionless coupling, $g\ll1$, while $\hat M$ and $\tilde M$ are mass scales, specified below. We assume that the function $f$ is invariant under $\phi_2 \to -\phi_2$, so that the full potential respects the $\mathbb Z_2$ symmetry. 

The first term in \eqref{potentialU} gives the potential for $\chi$ in the limit when its coupling to the ghost condensate field $\phi_2$ is turned off. For concreteness we choose it to be a Starobinsky-type potential \cite{Starobinsky:1979ty,Whitt:1984pd,Mukhanov:1990me}:
\be
W(\chi)=M_I^4 \left(1-e^{-\chi/M_I}\right)^2\;.
\ee
The other two terms in \eqref{potentialU} contain couplings to $\phi_2$ that have a finite range in the $\phi_2$ direction. Indeed, the second piece vanishes for $|\phi|>\phi_{\text{NEC}}$ by definition of the potential $V(\phi_2)$, while the third term is assumed to vanish for $|\phi|> \phi_f>\phi_{\text{NEC}}$ (see Fig.~\ref{fig:vphi}). 

\begin{figure}[t!]
\center
\includegraphics[height=3cm]{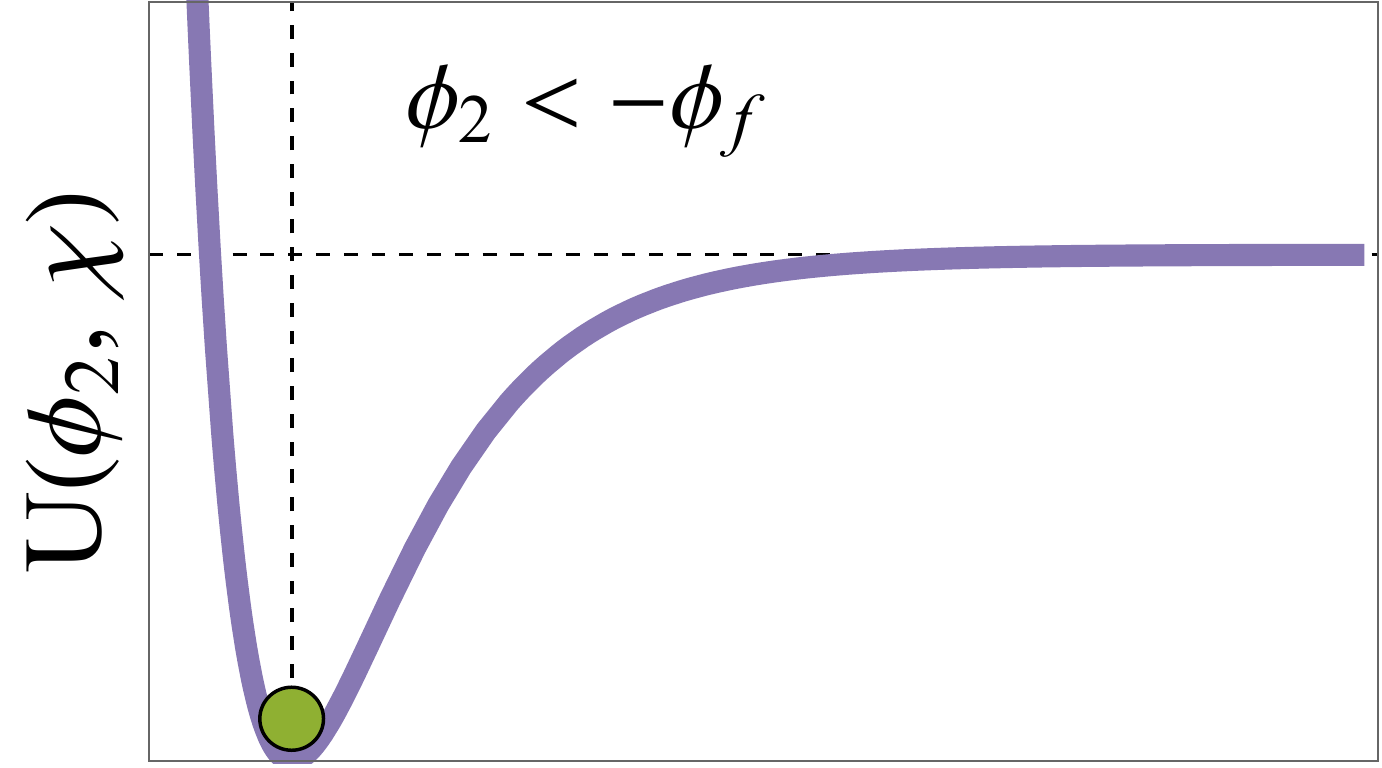}
\includegraphics[height=3cm]{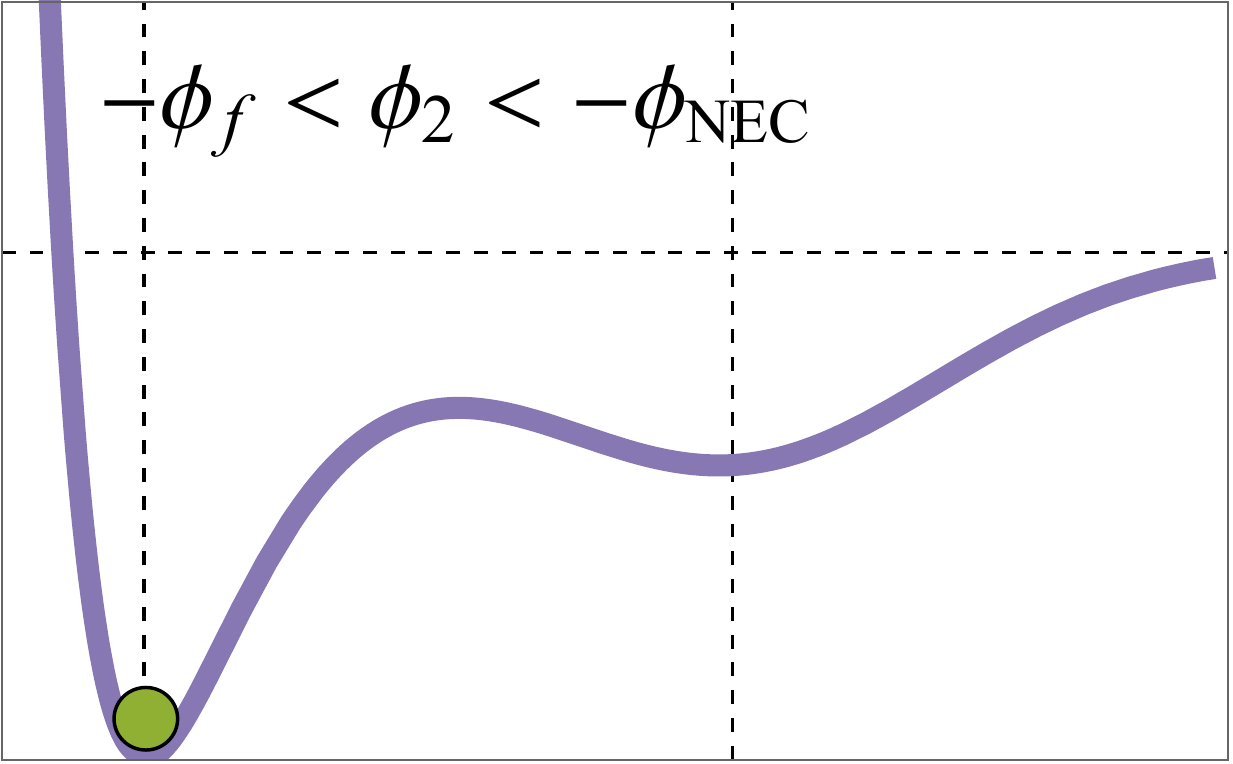}
\includegraphics[height=3cm]{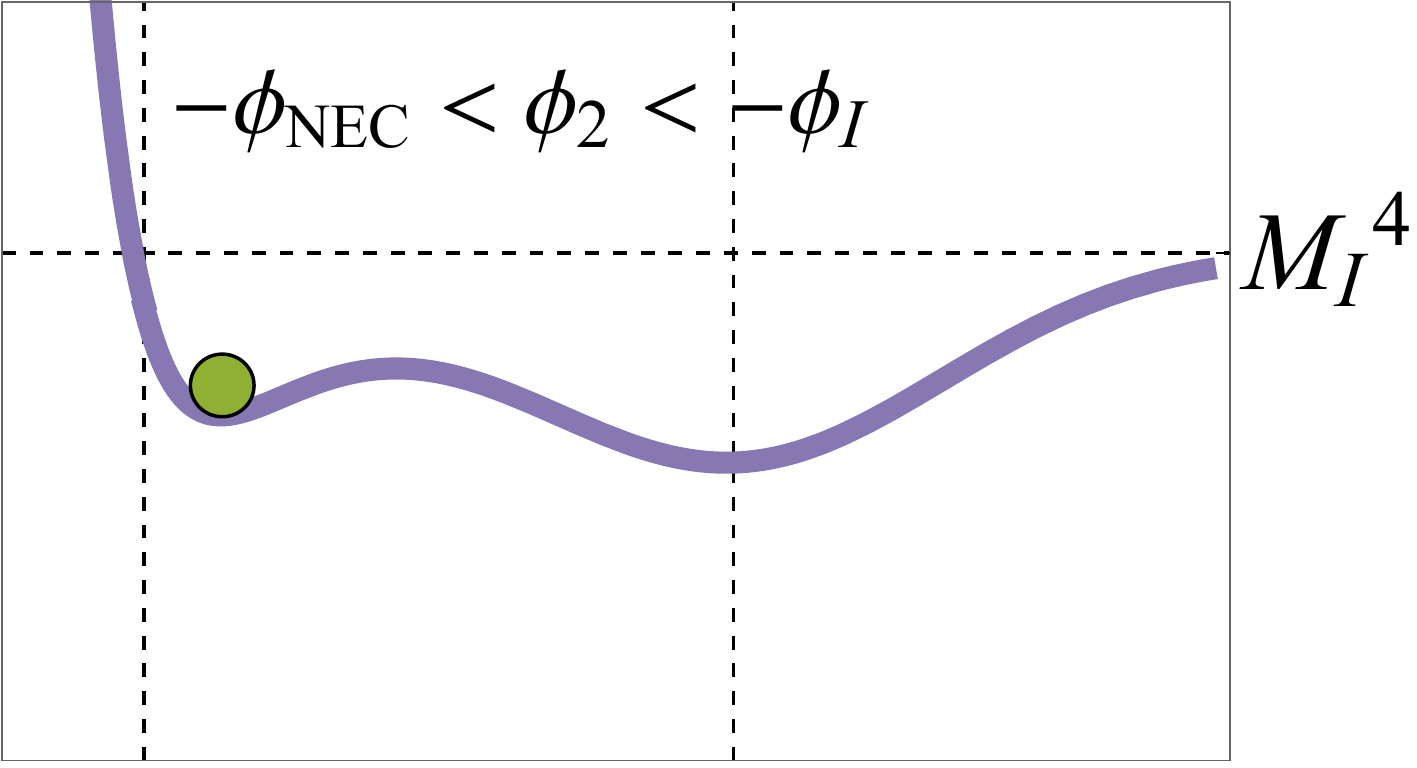}
\includegraphics[height=3.52cm]{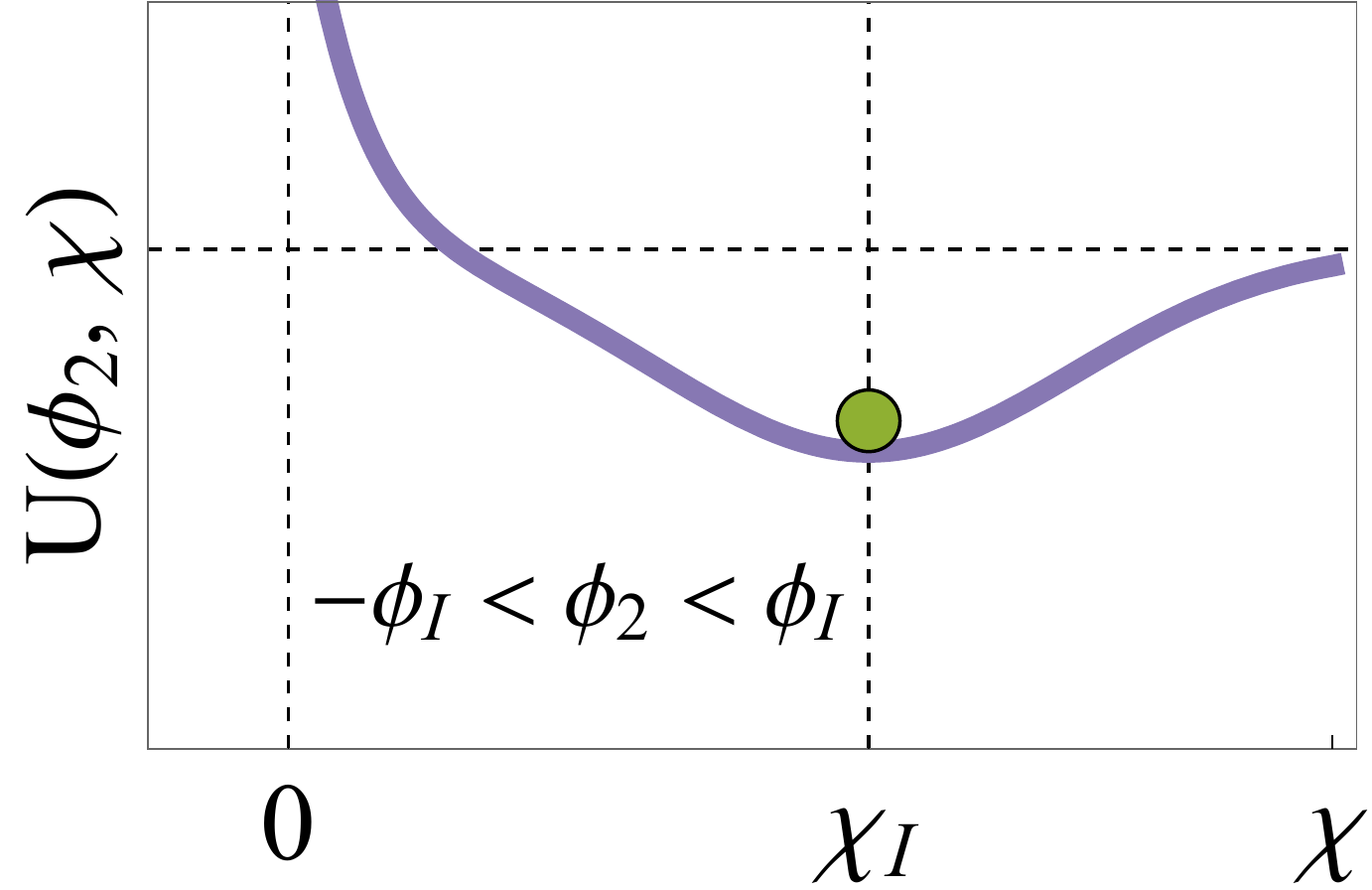} 
\includegraphics[height=3.52cm]{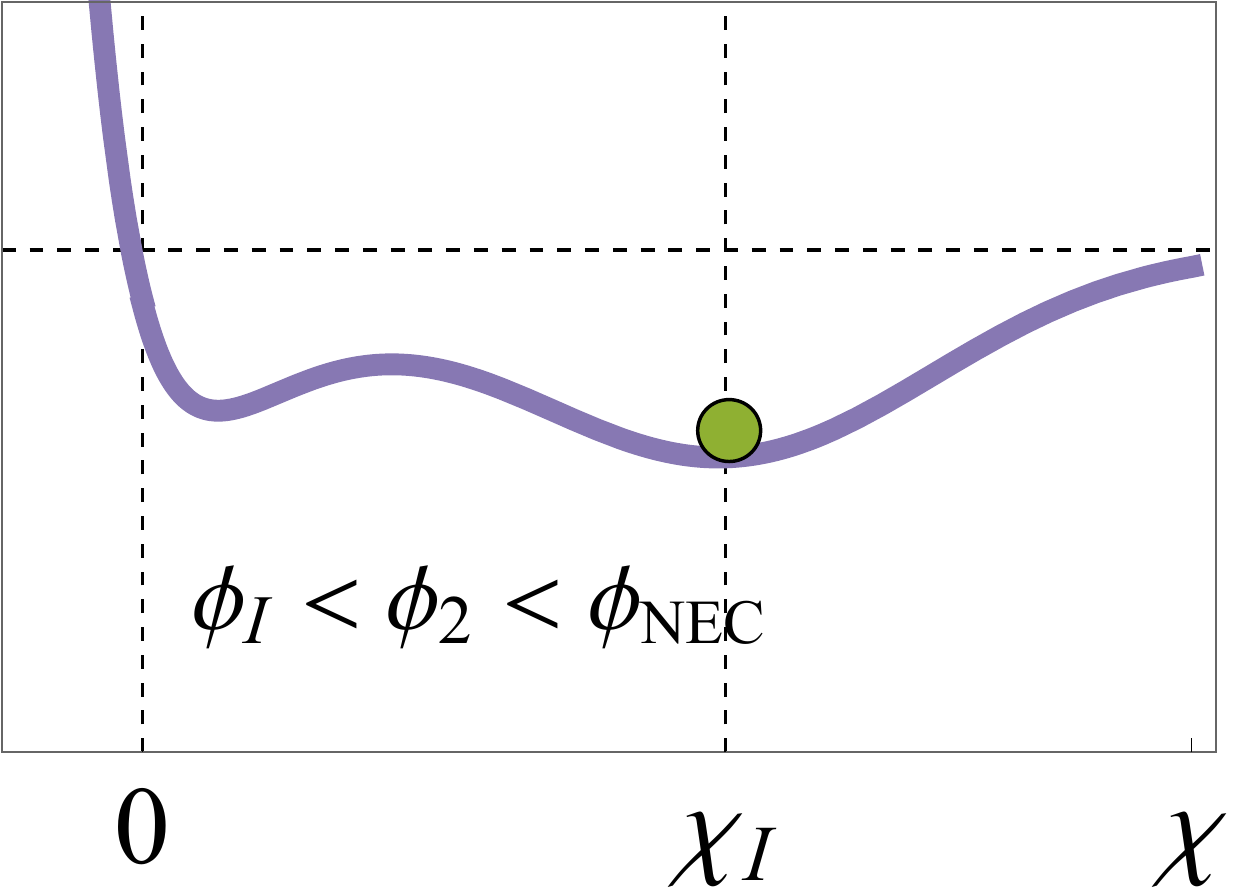}
\includegraphics[height=3.52cm]{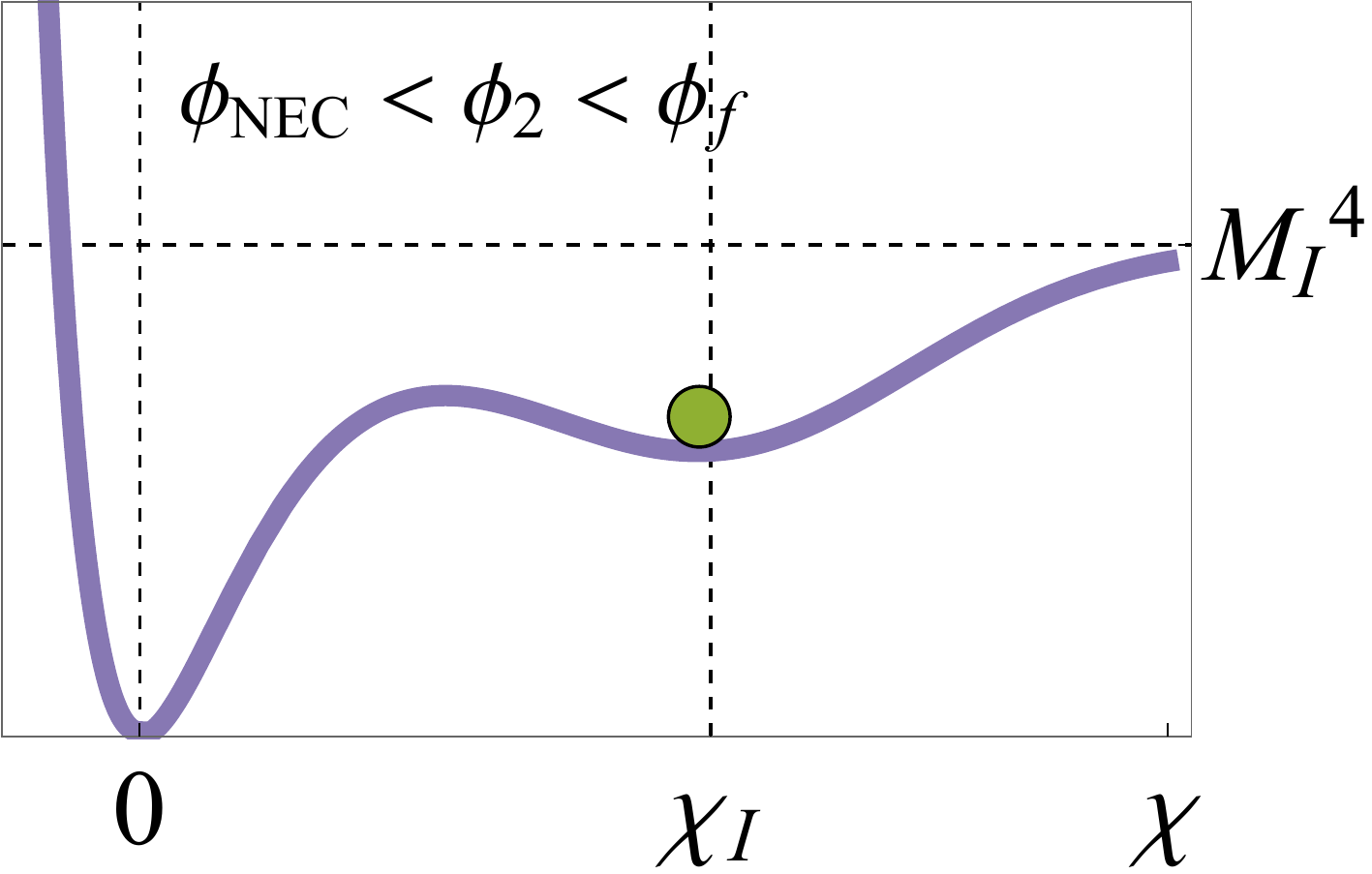}
\caption{A sketch of the potential $U(\phi_2,\chi)$ for different values of the NEC-violating field $\phi_2$.} 
\label{fig:potential}
\end{figure}

The potential $U(\phi_2,\chi)$ for $\phi_2<-\phi_f$ is shown in the top left panel of Fig.~\ref{fig:potential}. It becomes exponentially flat in the region where $\chi/ M_I\gg 1$ and has its only minimum at $\chi=0$, which we assume is the initial value of $\chi$. After the phase transition into a ghost condensate state with constant non-zero velocity, the field $\phi_2$ starts to move towards the point $\phi_2 = -\phi_f$ from the left. Upon crossing this point, the last term in the potential \eqref{potentialU} turns on, which results in the appearance of a second, shallow minimum at $\chi=\chi_I$ in the flat region where $\chi/M_I\gg 1$ and $U \sim M_I^4$ --- see the top middle panel of Fig. \ref{fig:potential}. Later on, when NEC violation sets in (\textit{i.e.} for $\phi_2>-\phi_{\text{NEC}}$), the initial minimum at $\chi =0$ starts to rise and gradually disappears (see the top right panel of Fig.~\ref{fig:potential}). This is due to the activation of the second term in \eqref{potentialU}, which significantly modifies the potential for $\chi$ for the field values $\chi/ M_I\ll 1$, but has exponentially suppressed effects in the flat region $\chi/ M_I\gg1$. On the other hand, the potential for the field $\phi_2$ for small field values $\chi/M_I\ll1$ (in particular, while the field $\chi$ sits in the minimum at $\chi=0$) remains almost unaffected by the coupling to $\chi$ and reduces to the linearly piecewise potential $V(\phi_2)$ shown in Fig.~\ref{fig:vphi}. As a result, while $\phi_2$ evolves on top of its potential ($-\phi_I<\phi_2<\phi_I$), $\chi$ rolls towards the second minimum at $\chi/M_I\gg 1$ and gets caught in it, acquiring potential energy of order 
\be
\rho_\chi\sim M_I^4~.
\ee
Starting from $\phi_2=0$, the shape of the potential changes in the opposite order, except that $\chi$ now sits in the false vacuum at $\chi=\chi_I$ when the ghost condensate reaches $\phi_2 = \phi_{\text{NEC}}$, as depicted on the lower right panel of Fig. \eqref{potentialU}. While $\chi$ is sitting in the false vacuum, the common potential $U(\phi_2,\chi)$ is nearly flat and the universe inflates with the inflationary Hubble rate of order $H^2_I\sim M_I^4/\mpl^2$. In the meantime, the inflaton field $\phi_2$ continues to roll and eventually reaches the point $\phi=\phi_{\text{NEC}}$ where the true minimum at $\chi=0$ has reappeared. Inflation ends when the ghost condensate field rolls past $\phi_2=\phi_f$, removing the second minimum in the potential \eqref{potentialU}.  Around that point $\chi$ starts rolling back towards the true minimum and oscillates around it --- eventually transferring energy to the standard model degrees of freedom through one of the conventional reheating mechanisms (see, \textit{e.g.},~\cite{Kofman:1997yn,Bassett:2005xm}). When $\chi$ reaches field values $\chi/M_I\ll1$ the inflationary period ends and one returns to the state with the present day cosmological constant. In this sense, the field $\chi$ plays a role analogous to the waterfall field that terminates the period of hybrid inflation~\cite{Linde:1993cn}. Notice that in our model the inflaton $\phi_2$ is a ghost condensate and this gives a peculiar phenomenology for primordial perturbations \cite{ArkaniHamed:2003uz}. In particular, the normalization of scalar perturbations will read $(H_I/M_2)^{5/4} \sim 10^{-5}$. Assuming $M_2^4 \sim \Lambda_0$, one gets $(H_I \mpl)^2 \sim (10\,\text{GeV})^4$: inflation occurs at very low energy, but it is still compatible with nucleosynthesis.

At this point we should comment on the $\phi_2$-dependence of the function $f(\frac{\chi-\chi_I}{\hat M},\frac{\phi_2}{\tilde M})$, that provides the mild modulation of the $\chi$ potential in the flat region. We will assume that $f$ is an order unity function of its dimensionless arguments. As remarked above, this function is symmetric under a sign flip for $\phi_2$. Moreover, given that $\phi_2$ varies over a huge distance in field space within the region of interest\footnote{This is a direct consequence of the fact that the slope of $V(\phi_2)$ is bound to be extremely small compared to its height, see Eq.~\eqref{lambda2upper2}. Assuming then, for example, that $M_I^4\sim \Lambda_\star$, that bound translates into $\phi_{\text{NEC}} > \mpl \(\mpl/H_0\)^{3/2}$.} ($-\phi_f <\phi_2<\phi_f$), one should make $f$ weakly dependent on this field. One way to achieve this 
is to assume that $f$ changes from zero to an order-one value within a region of width $\Delta\phi_2\sim \tilde M$ around $\phi_2\simeq-\phi_f$ and then stays constant all the way up to $\phi_2\simeq\phi_f$. The largest possible value of the $\phi_2$-derivative of $f$, therefore, is $\p_{\phi_2}f\,\lsim \,f/\tilde M$.

In what follows, we list several requirements that the potential in Eq. \eqref{potentialU} has to satisfy in order to provide a working realization of inflation and reheating in our scenario:

\begin{itemize}
	\item{The backreaction of the potential \eqref{potentialU} on the dynamics of $\phi_2$ should be negligible. This is guaranteed if the resulting correction to the velocity of $\phi_2$ is much smaller than its unperturbed value. The second term in \eqref{potentialU} clearly provides a sub-leading correction. The third term, on the other hand, gives
		\beq
		\dot \pi_2 \sim \frac{\p_{\phi_2} U}{H_I}\sim \frac{\hat M^4}{\tilde M \, H_I}\ll \dot \phi_2 = M_2^2 \;.
		\eeq
The above condition implies the following lower bound on $\tilde M$
		\beq\label{backreaction}
		\tilde M \gg \frac{\hat M^4}{M_2^2 \, H_I}\qquad\text{(no backreaction)}~.
		\eeq
		We note that this condition only concerns the backreaction on the dynamics of $\phi_2$ at the moment when expansion rate is $H\sim H_I$. One might suspect that the field $\phi_2$ experiences a stronger backreaction at the time of appearance of the modulation $f$ at $\phi\sim-\phi_f$ when $H\sim H_0$, leading to a much stronger constraint on $\tilde M$. However, the modulation $f$ is assumed to modify the shape of the total potential $U(\phi_2,\chi)$ only at values $\chi/M_I\gg1$. At the moment when $\phi\sim-\phi_f$, the field $\chi$ is at its minimum at $\chi=0$ and is assumed to have a negligible effect on $U(\phi_2,\chi)$.} 
		\item{In order for $\chi$ not to undergo significant quantum fluctuations in the false vacuum, its mass should exceed the inflationary Hubble rate
		\beq\label{mchi}
		m^2_{\chi}\sim \hat M^2 > H_I^2\qquad\text{(no fluctuations in }\chi\rm{)}~.
		\eeq
		This condition is equivalent to requiring that the slow-roll parameter $\eta=\mpl^2\,\partial_{\chi}^2U/U$ is large. }
		
		\item{As mentioned before, the field $\chi$ that rolls fast towards the true minimum of its potential after being released from the false minimum plays a role analogous to the waterfall field that terminates inflationary stage in the hybrid inflation scenario. During disappearance of the false minimum at $\chi=\chi_I$, the second derivative of the $\chi$ potential is expected to flip the sign from $\partial_\chi^2U>0$ to $\partial_\chi^2U<0$, so that the field $\chi$ subsequently rolls down to the true minimum at $\chi=0$. In such a scenario, the mass of $\chi$ crosses zero and thus violates the condition \eqref{mchi} during the release. When the mass of $\chi$ drops below the Hubble scale of inflation, its quantum perturbations contribute to the density perturbations generated during inflation together with the quantum fluctuations of $\phi_2$. For simplicity we choose to exclude such a possibility and put constraints that insure an effectively single-field inflationary stage\footnote{Notice that $\phi_1$ is light during inflation and therefore it gets quantum fluctuations. However the isocurvature perturbations produced in this way are completely negligible, as it is easy to check.}. In order for quantum fluctuations of $\chi$ to be negligible, the transition phase at $\phi\simeq\phi_f$ during which the inequality~\eqref{mchi} is violated should happen sufficiently quickly, preferably, within one Hubble time. We therefore require that at the end of inflation the field $\phi_2$ changes by order $\tilde M$ within one Hubble time and thus induces an order one change in the modulation $f$. This provides an upper bound on the scale $\tilde M$,}
		\beq\label{fasttransition}
		\(\Delta \phi_2\)_{H_I^{-1}} \simeq \frac{M_2^2}{H_I} \gtrsim \tilde M\qquad\text{(fast transition)}~.
		\eeq
		
	\end{itemize}

The bounds~\eqref{backreaction} and~\eqref{fasttransition} on the scale $\tilde M$ can be satisfied simultaneously only if the scale $\hat M$ of the modulation $f$ is smaller than $M_2$,
		\beq\label{hatm}
		\hat M \lesssim M_2 \lesssim \Lambda_0^{1/4} \sim 10^{-3}~\text{eV}\; .
		\eeq
		This leaves enough room to satisfy also the condition \eqref{mchi}, since $H_I/M_2 \sim 10^{-4}$.
		The condition \eqref{hatm} 
		implies that the scale of the locking potential is much smaller than the overall scale of the $\chi$ potential (that also sets the scale of inflation):
		\beq
		\frac{\hat M} {M_I} \lesssim \(\frac{\Lambda_0^{1/4}}{M_I} \) \sim 10^{-9} \cdot \( \frac{1~\text{MeV}}{M_I} \) \; .
		\eeq
		This confirms our assumption of $f$ being a mild modulation on top of the first two terms of the potential $U(\phi_2,\chi)$ in \eqref{potentialU}. Such a large hierarchy between the two scales can be explained as technically natural on the account of the overall flatness of the $\chi$ potential in this region.

\section{Conclusions and outlook}

The idea that the observed smallness of the cosmological constant may be a result of its dynamical relaxation is by no means new. In this work, we have put forward a concrete model, realizing this idea in a technically natural way. Our model shares some similarities with Abbott's thirty-year-old approach to the c.c.~problem \cite{Abbott:1984qf}, but also differs from it in several important ways. One difference is that our mechanism does not allow for a landscape of possible values of the observed vacuum energy. The latter is instead unambiguously fixed by the parameters in the Lagrangian. Importantly, the new sector is dominated by purely classical, rather than quantum, dynamics --- leaving no room for eternal inflation and the associated issues. Moreover, as pointed out by Abbott himself, the scenario of \cite{Abbott:1984qf} lacks a mechanism for producing significant energy density out of a relaxed low-curvature state of the universe, characterized by the Hubble rate of order $\sim 10^{-33}~\rm{eV}$. Specifying such a mechanism is necessary for connecting the latter state to inflation/Big Bang cosmology, and is thus an indispensable requirement for any model based on dynamical relaxation of the cosmological constant. Drawing on the last decade's progress in understanding theories that strongly violate the null energy condition, we have provided two explicit examples of how the inflationary universe may arise out of the post-relaxation low-curvature state.

Our model can be further explored along several interesting directions. We have not elaborated on the precise details of the phase transition that turns on the non-trivial dynamics for the NEC-violating sector. Neither have we attempted to find a UV completion of relaxation and/or NEC violation. Remaining agnostic about the latter, we have imposed the most stringent possible constraints on our model that results in the conservative upper bound of $\Lambda_\star ~\lsim~ (1 ~\rm{TeV})^4$ on the maximal magnitude of the relaxed cosmological constant. In principle, the dynamics of relaxation should be under complete control within a putative UV-extended theory of the ghost condensate. It would be interesting to see whether invoking such a UV completion, e.g. along the lines of Ref.~\cite{Ivanov:2014yla}, can lift the upper bound on $\Lambda_\star$ --- opening up a possibility to relax even higher values of the initial c.c.

Furthermore, our model directly links the cosmological constant problem to dark energy, and therefore provides a solid theoretical motivation for the experimental study of the current acceleration of the universe. On the theoretical side, it is important to explore the phenomenology of the late-time cosmology after the relaxation of the c.c. Depending on the precise scenario considered above, either one or both of the two (relaxing and NEC-violating) sectors of the theory contribute to the current dark energy, and it would be interesting to see whether this entails any sizeable observational consequences.
Can the traces of NEC-violation --- a crucial property of theories with c.c. relaxation --- be imprinted on today's universe in a way, amenable to observational verification? Would a putative UV extension of the theory introduce any novel, experimentally relevant features? We anticipate that the answers to these questions are model-dependent. For example, the precise phenomenology of dark energy --- e.g. the extent to which it differs from a c.c. --- would depend on which of the two scenarios of Secs.~\ref{sec:slownec} and \ref{sec:fastnec} is relevant. 
Exploring these matters has been largely left outside the scope of this work, and we plan to return to them in the future.    

\subsection*{Acknowledgements}
We would like to thank Alberto Nicolis, Riccardo Rattazzi, Sergey Sibiryakov, Gabriele Trevisan and Giovanni Villadoro for interesting discussions. We thank Guido D'Amico, Gregory Gabadadze, Justin Khoury, Mehrdad Mirbabayi and Valery Rubakov for valuable comments on the draft. D.P. thanks Riccardo Barbieri for valuable discussions on the cosmological relaxation of the electroweak scale. The work of D.P. is supported by the Swiss National Science Foundation under grant 200020-150060. The work of E.T. is supported in part by MIUR-FIRB grant RBFR12H1MW. 

\renewcommand{\em}{}
\bibliographystyle{utphys}
\addcontentsline{toc}{section}{References}
\bibliography{relaxingcc}

\end{document}